\documentclass[12pt,a4paper,aps]{article}
\usepackage{a4}
\usepackage{epsfig}
\usepackage[hmargin=2cm,vmargin=2cm,nohead]{geometry}

\usepackage{graphicx}
\usepackage{amssymb}
\usepackage{epstopdf}
\newif\ifpdf
  \ifx\pdfoutput\undefined
  \pdffalse
\else
  \pdfoutput=1
  \pdftrue
\fi

\ifpdf
  \DeclareGraphicsExtensions{.pdf,.jpg,.JPG,.jpeg,.gif}
\else
  \DeclareGraphicsExtensions{.ps,.eps,.eps.gz,.ps.gz}
\fi

\ifpdf
\usepackage[colorlinks=true, pdfstartview=FitV, linkcolor=blue, 
            citecolor=blue, urlcolor=blue]{hyperref}
\hypersetup{pdfstartview={Fit -32768},
  pdfpagemode=UseOutlines,
  plainpages=true,
  bookmarksnumbered=true,
  bookmarksopen=false
  }

\begin{document}

\title{A Regional Oil Extraction and Consumption Model. \\
Part II: Predicting the declines in regional oil consumption \\
} 


\author{
Michael Dittmar\thanks{e-mail:Michael.Dittmar13@gmail.com},\\
Institute of Particle Physics,\\ 
ETH, 8093 Zurich, Switzerland\\
\date{\today} 
}
\maketitle

\begin{abstract}
\noindent 
In part I of this analysis, the striking similarities of the declining oil production in the North Sea, Indonesia and Mexico were used to model the future maximum possible oil production per annum in all larger countries and regions of the planet from 2015 to 2050.\\

\noindent 
In part II, the oil export and oil consumption patterns, that were established in recent decades, are combined with the consequences of the forecast declines in regional oil  
production that were developed in part I of this analysis. The results are quantitative predictions of the maximum possible region-by-region oil consumption during the next 20 years. \\

\noindent 
The predictions indicate that several of the larger oil consuming and importing countries and regions will be confronted with the economic consequences of the onset of the world's 
final oil supply crisis as early as 2020. In particular, during the next few years a reduction of the average per capita oil consumption of about 5\%/year is predicted for most OECD countries in Western Europe, and slightly smaller reductions, about 2-3\%/year, is predicted for all other oil importing countries and regions. 
The consequences of the predicted oil supply crisis are thoroughly at odds with business-as-usual, 
never-ending-global-growth predictions of oil production and consumption. \\
   
\noindent 
Keywords: After the oil peak, regional oil production and consumption. 
\end{abstract}
\newpage

\section{Introduction}

Historical observers might eventually characterise the second half of the 20th century 
as a period of extraordinary economic growth and prosperity in many industrialised countries and for 
a significant fraction of  their citizens. In essentially all OECD countries, the resulting consumerist cultures 
evolved rapidly around an oil based transport system for people, raw materials and the products made with these materials. This system, from big ocean container ships, airplanes and trucks to millions of individual cars is founded on the availability of large amounts of oil and especially the refined oil products.  
This growth period also helped to heal many of the WW II related wounds in Western Europe, Japan and South Korea. Historians might perhaps argue, that today's globalised oil based mobility system 
actually began its rapid growth around 1935 and in the U.S. 

After WW II, some significant economic growth could also be observed in the Soviet Union and several countries in Eastern Europe. But, for whatever reasons, neither the individual car based mobility, 
nor the consumerism developed in any comparable way and this alternative model ended with an economic and political crash. 
 
In contrast, relatively little economic development was observed during this period in any of the former colonies in Africa, South East Asia and Latin America. Those newly independent countries and regions continued to extract their raw resources and export them to the richer former colonial powers and also to other rich industrial countries. In exchange for those resources, some consumer products were imported, but those products mostly served the newly established elites and a developing global tourist industry. 

In the OECD countries in particular, the simultaneity of this economic growth with the rapid 
growth in the consumption of oil, which served as the energy resource for the individual
car mobility and the transport of consumer goods, is remarkable \cite{EIAoilhistory}.
For example oil equivalent liquid energy consumption in the U.S. grew between 1935 to 1950 and 1950 to 1973 by factors of about 2.2 and 2.8 respectively, corresponding to an average growth of about 5\%/year. From the mid-1960s to 1972 similar oil consumption growth rates were observed in the larger European Countries and in Japan. For example in just seven years, between 1965 and 1972, the oil consumption in Germany, France, Italy and the UK grew by factors of 1.8,  2.1, 1.9 and 1.5 respectively and by an even 
larger factor of 2.7 in Japan. 

This growth period ended rather suddenly, with what is often called the first and second oil (price) shocks, 
in 1973 and 1979. During the subsequent economic crises, oil consumption declined, especially in the oil importing industrial OECD countries. 
These crises led to (1) the phasing out of oil usage for the production of electric energy and its replacement by nuclear and natural gas as energy sources and (2) the transition from oil-based heating systems to gas and electric heating systems combined with better insulation techniques.   

Despite this overall oil consumption decline, the oil demand for individual car mobility, for the transport of consumer goods and for air traffic continued to grow. As a result, oil consumption in the OECD countries as a whole started to grow again after 1983 and continued to grow until oil prices once more 
increased strongly around 2005. Typical oil consumption growth rates during that time were about 1-2\%/year in the U.S. and Japan, while the consumption in many of the larger countries in Western Europe was roughly constant. After 2005, the higher oil prices led to the end of oil consumption growth. This was followed by a decline in oil consumption in the OECD countries
during and after the 
economic crisis in 2008/09,  and most of those countries have not yet returned to their 2005 levels of oil 
consumption.   
   
During the last decade of the 20th century and the beginning of the 
21st century, a similar, or perhaps even faster economic growth period than the one seen in the OECD countries after WW II, was observed in China and India and several other former colonies.  
And, like the last, this economic growth period coincided with rapidly growing oil consumption. 
For example the oil consumption in China and India grew 
between 1994 and 2015 by a factor of almost 4 and 2.6, corresponding to an average annual growth rate of about 7\% and 5\%, respectively.  And in contrast to the OECD countries, it seems that the growing appetite for oil in China and India was, at least up to that point, little affected by the periods of high oil prices between 2005 and 2014. 

The per capita oil consumption in many Sub-Saharan countries, no matter whether they 
are substantial oil producers or not, is essentially still negligible. 
Taking the situation in Nigeria, with its huge population of about 186 million people, 
as an example, one finds oil production of up to 2.5 mbd, corresponding to about 
730 liter per capita and year. But almost 90\% of this oil is exported to Western Europe, Asia and the U.S.,
which explains to a large extent why significant economic growth has been impossible in Nigeria.
Other oil-exporting former colonies in the region have faced the same issue.   

Table 1 summarises more details about official conventional oil reserves, production and consumption in some of the larger consumer, producer and exporter countries. The corresponding data are taken from 
the CIA World Fact Book \cite{CIAfact}, the EIA \cite{EIA} and from the Statistical World Energy Review from BP
\cite{BPstat}. 
  
{\tiny
\begin{table}[h]
\begin{center}
\begin{tabular}{|c|c|c|c|c|c|}
\hline
Country & population & population  & reserves         & crude production      & oil* consumption   \\
             &    [Million]   & growth [\%] & [liter/capita]   & [liter/capita/year] & [liter/capita/year] \\
\hline
World      & 7400         & 1.1             &   28 000 &   720  & 750 \\
\hline
U.S.     &   324 &  0.8  &     23 000  &  2 300  & 3 500  \\  
Japan     & 127 &  -0.2  &       0         &  0       & 1 900  \\
Germany &   81 &  -0.15  &     200   &    35   & 1 700  \\
EU           &   515 &  0.25  &    1 600 &  170  &  1 400    \\
\hline
Brazil   &   206 &  0.75  &      10 000   &   700  &   890  \\
China  &  1 373 &  0.4  &         3 000  &    180  &   510  \\
\hline 
Egypt   &  95 &    2.5  &            5900    &  440  & 500 \\
India    &  1 267 &  1.2  &          600     &     40  & 190  \\
\hline
\hline
Saudi Arabia  & 28.2 &  1.45   &       1 500 000 & 24 700 & 8 000 \\
Kuwait  & 4.2  & 1.5                  &   3 840 000    & 43 000     & 7 300 \\  
UAE     & 9.3  &  2.5    &  1 670 000     & 24 300    & 5 600 \\
Quatar & 2.3  &  2.65    &  1 740 000     &  48 000   & 8 200  \\
Iran   & 82.8 & 1.2          &     304 000     & 2 700      & 1 400 \\  
Iraq   & 38.1 & 2.9          &     595 000     & 6 100      & 1 200 \\  
\hline
Norway       &     5.3 & 1.1      &     240 000   &  21 300     & 2 600   \\
Russia        &  142.4 & -0.1 &        114 000   &   4 500      & 1 300  \\
Mexico        &  123.2 & 1.15 &        10 300   &   1 200      &     900  \\
\hline
Algeria         &  40.3 & 1.8     &         48 000      &   2 300     &  610 \\
Angola        &  20.2   &  2.7    &        93 000       & 5 200      &  380  \\  
Nigeria        &  186.1 &  2.45   &       32 000       &   730      &   90 \\
\hline
Venezuela  &  30.9 & 1.3     &         404 000     & 4 900     &  1300 \\
Ecuador     &  16.1  &  1.3    &          79 000       & 2 000      &    910  \\  

\hline
\end{tabular}\vspace{0.1cm}
\caption{2015 data for some larger oil importing and oil exporting countries. The official conventional oil reserves, as well as production and consumption data are given in liter/capita and liter/capita/year respectively. The  country data are taken from the CIA world fact book \cite{CIAfact}, the EIA \cite{EIA} (the U.S. Energy 
Information Agency), or   
the Statistical World Energy Review from BP \cite{BPstat}.  The estimated remaining conventional crude oil per capita is based on official claims regarding remaining ``assured reserves". 
The production numbers include all crude oil equivalent fossil fuels, and ``oil*" consumption, which is globally about 15\% larger than production, 
includes not only all oil equivalent liquids but also ``refinery gains".} 
\end{center}
\end{table}
}
Ignoring for now the uncertainties of official country crude oil reserves, the data presented in Table 1 show 
that crude oil is definitely a finite and very unequally distributed global resource. 
The last three columns also indicate that the people in oil exporting countries 
are not automatically benefitting from their countries' resource wealth and that, at least outside of the oil rich 
OPEC Middle East region, it is currently more profitable to import and use the oil instead of 
making the economy function on the income from crude oil exports.  
A more careful comparison shows a huge inequality in the average per capita oil consumption in the richer industrial countries and oil exporting countries especially in Africa. 
The unjustness of this inequality is especially obvious when reading daily news about the civil-war-like violence in Nigeria and Angola and their rapidly growing very young and poor population. If one accepts that their remaining oil reserves are indeed finite, future industrial and economic development possibilities in those African and South-American countries look less and less promising by the day.

In contrast to the poor countries just mentioned, the rich, less populous, oil-exporting OPEC countries  
in the Middle East - Saudi Arabia, Kuwait, United Arab Emirates and Qatar -, with a combined population of 
about 30 Million (plus about 15 million of immigrant workers) have a per capita oil consumption 
that is not only higher than that of the U.S. but is considerably more than half again as much. 
However, it should be noted that oil (and natural gas) in these countries, in absence of alternatives like coal and hydropower, is used to produce electricity. This explains to some extent their very high per capita oil consumption.  

In the following, the oil export and oil consumption patterns that were established in recent  
decades are examined in light of the regional declines in oil production that were predicted
in part I \cite{Part1paper}, of this analysis. The results are quantitative predictions of maximum possible 
oil consumption region by region.

The starting point, section 2, will be an analysis of oil import, export, and consumption 
patterns as they have evolved over the past decade. The next decades expected 
production trends in currently importing and exporting regions, as obtained from Part I of 
this analysis, are then used, in section 3, to examine the impact that declining production
will have on exports. In part 4, the estimated declines in regional oil consumption during 
the next decades are contrasted with the expectations from other, mostly economically
inspired, "demand forecasting models" of large organisations like the IEA. The paper is 
completed with a discussion of some of the consequences of the expected declines
in oil consumption in different parts of the planet, some richer, some poorer.

\section{Regional oil trade and consumption trends since 2005}

Before describing the regional oil trade trends, it might be relevant to notice the 
reported differences between the production of crude oil and other fossil hydrocarbon equivalent liquids
and the consumption of the products from the refinery of crude oil and other energy liquids. 
For example, the consumption of all oil equivalent and refined products in 2015 is reported 
as being globally about 3.3 mbd (or 3.5\%) larger than the production. 
This difference is only partially explained with the contributions of 1.8 mbd attributed to biomass fuels 
and 2-2.5 mbd of fuels counted as refinery gains. In addition, some losses, usually unquantified, occur during the crude oil transport and in the refinery process were non liquids, like asphalt, are produced. 

It might also be interesting to analyse the current and future refinery problems and the 
quality of the remaining exploitable regional crude oil, e.g. how much of it is either light or heavy and how much sulphur it contains and how this impacts the operations of crude refineries and the environment. 

Furthermore, as crude oil is essentially not usable directly, the regional trade patterns and oil usage are 
at least partially controlled by the owners of refineries. This is especially evident when one 
looks at the numbers for the U.S. According to the BP report, during 2015 the U.S. imported 7.35 mbd of crude oil and 2.05 mbd of products, but exported 0.5 mbd of crude oil and 4.1 mbd of products. This corresponds to a net import of 6.85 mbd of crude oil and a net export of 2.05 mbd of refined products.
In more detail one finds that the crude oil exports from Central and South America were about 1.6 mbd.
This almost matched the imports of refined products of about 1.2 mbd from the U.S. back to Central and South America. Similar numbers are found for the trade of crude oil and products between Western Europe   
and North and West Africa. Those two regions exported about 1 mbd and 1.7 mbd of crude oil 
to Western Europe and imported almost 1 mbd of refined products. Knowing that the per capita oil product consumption in those poorer crude oil producer countries in Africa and South America 
is small, one might conclude that a significant economic development of the poorer producer countries requires that they take control of the refining process. Such changes would obviously reduce 
their oil exports to richer countries. 

A more detailed analysis of the possible evolution of the future local refining industry should reveal many more interesting details about the rich oil consumer countries and the poor crude oil producer countries. However, such details are beyond the scope of the analysis presented in this paper. 

In any case, the crude quality and refinery problems are assumed to be smaller than the uncertainties related to the often unpredictable consequences of regional oil consumption that are caused by perturbations of the global economical and political situation. 

Therefore, the interregional and intercontinental crude oil and product import/export trade will be approximated herein as the difference between the production and consumption of crude oil equivalents in the 
exporting countries as provided by BP's yearly ``Statistical-Review-of-World-Energy" \cite{BPstat}. 
 
As a result, one estimates the net crude oil trade between different countries and regions
to be about 40 mbd in 2015. And subtracting the intracontinental crude transport in different regions and especially in North America, about 2.1 mbd and 0.7 mbd were exported in 2015  from Canada and Mexico, to the U.S. and about 1.7 mbd from Norway to the EU, 
one estimates that the net intercontinental oil equivalent trade was about 36 mbd. 

About 60\% (20.5 mbd) of those 36 mbd originated from the OPEC Middle East Gulf countries,   
and another 10 mbd comes from three FSU countries - Russia, Kazakhstan and Azerbaijan.  
In comparison the exports from North and Sub-Saharan Africa and South America are 
much smaller, just 4.5 mbd and 1.5 mbd respectively.

Before discussing the consequences of the declines in oil production that are expected 
in all regions except the OPEC Middle East countries, some numbers regarding the 
net export and imports in 2015 and the evolution of the 
intercontinental oil trade and consumption between 2005 and 2015 are quantified in Table 2 and 3,
which are discussed in more detail in section 2.1 and 2.2. 

{\tiny
\begin{table}[h]
\begin{center}
\begin{tabular}{|c|c|c|c|c|}
\hline
Exporting to / Importing from & China &  other Asia Pacific & USA &  Western Europe \\
country or region   &    [mbd]   & [mbd]                & [mbd]   & [mbd] \\
\hline
Middle East       &  3.4    &   11.7               & 1.6        &  1.8       \\
FSU region       &   1.0    &   1.2                 &  0.4       &  6.2      \\
Africa                &  1.3     &   0.8                 &  0.3       & 1.8       \\
South America &  0.9     &  0.8                  & 0.4        & 0.2        \\
\hline
\end{tabular}\vspace{0.1cm}

\end{center}

\caption{Quantitative numbers for the net trade of crude oil and oil equivalent products from exporting to importing regions in 2015, as obtained from the BP world energy review 2016 \cite{BPstat}.
It should be noted that the trade of crude oil and products, as reported by BP, matches only 
approximately the difference between consumption and production in different regions. 
This might be explained in part by incomplete or flawed data on ``refinery gains" and  ``oil equivalent liquids"  
and by difficulties encountered when attempting to specify the origin of oil that is used for transports 
freight by ships and planes from and to different regions. 
} 
\end{table}
}

\subsection{Oil equivalent liquid exports and imports in 2015}

\subsubsection{Exporting regions}
~ 

{\bf OPEC Middle East Gulf countries}\\

\noindent 
For decades, the OPEC Middle East Gulf countries have been by far the dominant oil producing and exporting region of the planet. Their combined oil production, according to the BP report, was 28.8 mbd in 2015.
In comparison the production in the other Middle East countries is given as 1.3 mbd which is 
which is slightly smaller than their consumption.     

After subtracting the internal consumption in the Middle East countries, which includes some 
intraregional trade, the total net exports from the OPEC Middle East countries can be determined. 
According to the annual report from BP \cite{BPstat}, the consumption in the Middle East region increased from 6.6 mbd (2005) to 9.6 mbd  in the 2016 report (the number has been reduced to 9.3 mbd
in the 2017 report). Taking the BP 2016 numbers the net exports from the Gulf countries increased 
from 18.9 mbd in 2005 to about 
20.5 mbd in 2015, about 2/3 of their total production, essentially all of which was transported by tankers through the Persian Gulf. 

Saudi Arabia was the dominant supplier exporting about 8 mbd. 
The other 13 mbd came from Iraq, Iran, Kuwait, the Arab Emirates and Qatar. 
During 2015 about 16 mbd (almost 80\%) of the total net exports, were transported to Asian countries.  

Looking at the globe, one finds that this is easily explained by the fact that shipping lines to Asia are about a factor of 4 shorter than the lines to the U.S. and Western Europe\footnote{Oil to Western Europe has to be transported around the African continent, as the Suez Canal is too narrow for the big oil tankers.}. \\

{\bf Former Soviet Union countries}\\
  
\noindent 
The second large oil exporting region, with a total production of about 13.5 mbd in 2015, is formed 
by three former Soviet Union countries (FSU): Russia (10.98 mbd), Kazakhstan (1.67 mbd) and 
Azerbaijan(0.84 mbd). After subtracting the internal consumption, Russia,  Kazakhstan 
and Azerbaijan exported about 8 mbd, 1.4 mbd and 0.75 mbd respectively. About 6 mbd (or about 60\%) 
of this oil was exported, mostly through pipelines, to Western Europe. The remaining oil exports were 
sent to China, Japan and other Asian countries.  \\

{\bf Africa and South America}\\

\noindent 
The other oil exporting regions are Africa (North Africa and Sub-Saharan Africa) and South America. 
Africa produced 8.4 mbd in 2015 and, with very low internal consumption per capita,  
exported about 4.5 mbd to other continents. 
Most of those exports came from just 4 countries, about 1.5 mbd from Algeria and Libya and another 2.5 mbd from Angola and Nigeria. And most of the exports were sent 
to Western Europe (1.8 mbd), China (1.2 mbd) and 
India (0.7 mbd). In 2015, only about 0.3 mbd and 0.2 mbd were shipped to the U.S. and South America, respectively.

For South America one observes that in addition to crude oil production, which was about 7.7 mbd, 
more than 1 mbd of oil equivalent was produced from biomass and liquified gas. 
With a total consumption of 7.1 mbd, one finds that most of the produced oil equivalent in 2015 was consumed within the continent, and that net exports in 2015 were at most around 1.6 mbd, of which about     
0.9 mbd were sent to China and about 0.4 mbd to the U.S.\footnote{Taking the small imports 
from Africa and the Middle East into account, a discrepancy between the exports 
reported by BP and the difference between production and consumptions of about 0.2 mbd remains.} \\

\subsubsection{Importing regions}
~ 

{\bf Asia Pacific}\\

\noindent 
Regarding importers, the most populous are the Asian Pacific countries which, taken together, 
constitute about 50\% of the global population.  
Total oil consumption in 2015 in these countries is reported to be 32 mbd, and about 
24 mbd (75\%) of this is imported. 
Around 16 mbd (about 2/3 of the imports) come from the OPEC Middle East countries. 
The rest of the imports come from the FSU countries, from Africa, and from South America.   \\ 

{\bf Western Europe}\\

\noindent 
Western Europe, with about 7\% of the global population, has only two 
significant oil producers - the UK and Norway - both of which are suffering steep production declines. 
Only Norway is still exporting oil, and its exports are only 1.7 mbd.
   
In 2015, total oil equivalent liquid consumption in Western Europe was 
about 13.2 mbd, about 14\% of the world's total. 
To satisfy that consumption, net oil equivalent imports were 9.8 mbd. 
About 5 mbd came from Russia, and from Kazakhstan and Azerbaijan together came another 1 mbd.  
An additional 1.8 mbd came from Africa, and from the OPEC Middle East countries came another 
1.9 mbd. \\

{\bf North America (U.S.)}\\

\noindent 
North America's total oil equivalent liquid consumption, including biomass, was 23.6 mbd in 2015, 
about 19.6 mbd of which were produced within the region. Only 6\% of the world's people live 
in North America, but they consume about 24\% of the world's oil production. 
North America is, therefore, by far the largest oil per capita consuming continent on the planet. 
However, of the three North American countries, 
only the U.S. is a net importer: the other two countries, 
Canada and Mexico, exported about 2.1 mbd and 0.7 mbd respectively to the U.S. 
It should be noted that the per capita consumption in Mexico, about 900 liter
per year, is about a factor of 4 smaller than the per capita consumption in the U.S. and Canada.  

Intercontinental oil imports in 2015 to the U.S. come from  
the OPEC Middle East countries (1.6 mbd),  South America (0.6 mbd) and from Western Africa (0.3 mbd). 

All of the above imports should be compared to total U.S. production of oil equivalent liquids, 
which are reported by BP to have been 12.7 mbd in 2015. 
This production consisted of 8.8 mbd crude oil and condensates (about 5 mbd from 
shale or tight oil), a bit less than 3 mbd of NGL (natural gas liquids) 
and about 1.1 mbd from refinery gains. 

\subsection{Oil consumption and oil trading trends during the last decade}

As can be seen in Table 3, the regional and larger country imports and exports 
from different parts of the world changed considerable between 2005 and 2015. 

The combined intercontinental exports from the Middle East, the 
FSU countries, African and South-America were roughly the same in 2005 and 2015.
However, as can be seen in the following subsections, exports to most OECD countries declined, 
while more oil was exported to China and India - oil that came essentially from all exporting regions.

{\tiny
\begin{table}[h]
\begin{center}
\begin{tabular}{|c|c|c|c|c|}
\hline
Country/ & consumption 2005 &  imports 2005 &  consumption 2015 & imports 2015 \\
Region   &    [mbd]   & [mbd] & [mbd]   & [mbd] \\
\hline
U.S.  &  20.8    &   13.9 (2 + 1.4)  & 19.4  &   6.7 (2.1+0.7)    \\
EU + Nor + CH & 15.6     &   9.9   & 13.2  &  9.8      \\
Japan & 5.3     &  5.3    & 4.2  &  4.2      \\
China &  7.2    &  3.6    & 12.3  &  8.0      \\
India &  2.6    &   1.9   & 4.2  &  3.3      \\
Other Asia/Pacific & 9.5 & 5.8   & 11.8  &  8.6      \\
combined & 61     & 40.4    & 65  &  40.6      \\
\hline
Country/ & consumption 2005 &  exports 2005 &  consumption 2015 & exports 2015 \\
Region   &    [mbd]   & [mbd] & [mbd]   & [mbd] \\
\hline
Middle East &   6.6   & 18.9 & 9.6   & 20.5 \\ 
Russia & 2.6 & 7.0 & 3.1 & 7.9 \\
other FSU & 1.0 & 1.2 & 1.0 & 1.9 \\ 
North Africa &   0.9   & 3.6 & 1.3   & 1.5 \\
other Africa &    2.1   & 3.2 & 2.6   & 3.0 \\
South America &   5.3   & 2.0 & 7.1   & 0.6 \\   
Combined        & 18.6   & 36  & 25  & 35.4 \\ 
\hline
\end{tabular}\vspace{0.1cm}

\end{center}
\caption{Net trade and consumption figures from 2005 and 2015, according to the 2016 BP world energy review, \cite{BPstat} are given for the larger oil importing and oil exporting regions/countries.
The consumption numbers include all oil equivalent liquids and refinery gains,
which are globally about 15\% larger than the production numbers. The U.S. 
numbers in parentheses are the net intracontinental imports from Canada and Mexico respectively.} 
\end{table}
}

\subsubsection{Oil consumption and exports from the OPEC Middle East} 

From 2005 to 2015, the oil equivalent production of the six OPEC Middle countries increased 
from 23.7 mbd to 28.9 mbd. However since the internal consumption in the entire Middle East region increased by about 3 mbd, and since wars in the region reduced oil production in the non-OPEC countries, net exports from the region changed only slightly from 18.9 mbd to 20.5 mbd, and most of this additional oil was exported to China and India.

\subsubsection{Oil consumption and exports from Russia, Kazakhstan and Azerbaijan} 

Between 2005 and 2015 the combined production from Russia, Kazakhstan and Azerbaijan increased 
by 2.1 mbd. Since their internal consumption grew only from 2.9 to 3.5 mbd, and the production and consumption in the other FSU countries decreased during that period from 0.5 mbd to 0.4 mbd  
and 0.7 mbd to 0.6 mbd respectively, the interregional exports 
from the FSU region increased in from 8.2 mbd in 2005 to 9.8 mbd in 2015.  
However, as the additional oil production came dominantly from 
new fields in Eastern Siberia and Eastern Kazakhstan, this oil flows through 
new pipelines mainly to China. As a result, exports to China increased significantly, while 
the exports to Western Europe were only increased by 0.4 mbd, from 5.8 mbd to 6.2 mbd and did not 
compensate for the steeply declining oil production in Western Europe.  
 
\subsubsection{Oil consumption and exports from Africa and South America}

Geographical distances and historical colonial relations, explain why the largest fractions of oil exports 
from North- and Sub-Saharan Africa and from South America are exported to Western Europe and the U.S. respectively. 

Despite the very small per capita oil consumption in essentially all countries in Africa, 
the consumption has nevertheless increased from 2.9 mbd in 2005 to 3.9 mbd in 2015. 
And since production in North Africa and Sub-Saharan Africa decreased from 9.8 mbd to 8.4 mbd, exports in the region declined from 6.9 mbd to 4.5 mbd. As a result, net exports to Western Europe and the U.S. declined from 2.4 mbd and 2.5 mbd respectively in 2005 to 
1.8 mbd and 0.3 mbd respectively in 2015. Surprisingly, exports to China and India increased from 
0.6 mbd to 1.7 mbd.  

For South America, the total consumption of all liquids, including biomass and derivatives from natural gas, increased between 2005 and 2015 from 5.3 mbd to 7.1 mbd. As the crude oil and NGL production increased only slightly from 7.3 mbd to 7.7 mbd, net exports declined from 2.5 mbd in 2005 to 1.6 mbd in 2015. 

The changes in the oil trade with the U.S. are especially interesting. 
In 2005, about 2.9 mbd of crude oil were exported to the U.S. and 0.3 mbd of the refined 
products were sent back to South America. But in 2015, only 1.6 mbd 
were exported to the U.S. and 1.2 mbd of refined products were sent back to South America.   
Exports to China and India were almost negligible in 2005, but are reported to have been 1.4 mbd during 2015.
 
\subsubsection{Oil consumption trends in the importing regions}

In many of the OECD countries, oil consumption declined considerably during the high-oil-price 
period which started around 2005 and the decline was especially significant after 
the global economic crisis in the fall of 2008. 
Consumption in the U.S., the EU and Japan declined by 1.3 mbd (7\%), 2.3 mbd 
(14\%) and 1.2 mbd (22\%) respectively between 2005 and 2015. 

The decline of consumption observed in many Western European OECD countries 
coincides with the decline of Western Europe's oil equivalent production, especially from the North Sea. 
European production decreased from more than 6.7 mbd in 2002  
to 5.7 mbd in 2005 and to about 3.5 mbd in 2015. 

Until 2007, Europe's consumption held fairly constant, at about 15.5 mbd, and 
Europe's production decline was compensated by slightly increased imports, mostly from Russia.
After 2008, however, consumption decreased 
significantly, especially in the Southern European crisis countries, and reached about 13.3 mbd in 2015. 
With production declining 
3.2 mbd from 2002, and consumption declining only 2.2 mbd from 2002, imports had to 
increase by 1 mbd. In total, the European OECD countries imported about 10 mbd in 2015.

The U.S. is the only net oil equivalent importing country in North America. 
Due to the rapid increase of unconventional tight oil production in the last few years, 
from essentially zero in 2005 to  
to almost 5 mbd in 2015, the U.S. managed to reduce its overall net oil imports from 12.4 mbd in 2005 to 
4.9 mbd (2015). Imports from Canada and Mexico together remained roughly constant at 3.5 mbd, 
but more was imported from Canada and less from Mexico as that country's production declined.
Imports from Mexico fell from 1.4 mbd in 2005 to 0.7 mbd in 2015.  

Given the increase in U.S. production and steady intracontinental imports, and given as well 
that overall oil consumption in the U.S. declined by about 1.4 mbd (7\%) between 2005 and 
2015 due to the high price of oil and the aftershocks of the 2008-2009 economic crisis, it is 
not surprising that intercontinental imports to the U.S. decreased. They fell from about 9 mbd 
in 2005 to less than 2 mbd in 2015. 

Especially remarkable are China's (including Hongkong) and India's increased oil consumption during this period. 
While China's and India's production increased only from 3.6 mbd and 0.74 mbd in 2005 to 4.3 mbd and 0.88 mbd in 2015 the consumption increased, despite the steeply rising price of oil from 2005 to 2013, 
from 7.1 mbd  and 2.6 mbd in 2005 to 12.3 and 4.2 mbd in 2015 respectively.
It follows that China's and India's imports grew from 3.5 mbd  and 1.9 mbd in 2005 to 8.0 mbd and 3.3 mbd in 2015. 

According to the BP 2006 and 2016 reports \cite{BPstat}, China's crude oil imports from the Middle East OPEC countries increased from 1.4 mbd in 2005 to 3.4 mbd in 2015. 
Other crude oil imports to China in 2015 came from Russia and Kazakhstan (1 mbd), 
from Africa (1.3 mbd) from South America (0.9 mbd). 
The trade of products, as specified in the BP report, adds roughly another 1 mbd of refinery products from different regions and from other Asia Pacific countries to China's imports.  Thus, not all of the net Chinese imports of 8 mbd in 2015 can be traced from the BP report. Some of the missing oil equivalent liquids 
might perhaps be explained from refinery gains.
Accordingly one finds that India imported 2.3 mbd from the Middle East, 0.6 mbd from South America and
0.7 mbd from Africa. 

Studying the BP reports one finds that total oil exports from the Middle East to 
Western Europe and the U.S. decreased from 3.1 mbd and 2.3 mbd respectively 
in 2005 to 2.4 mbd and 1.5 mbd in 2015. And, as the export to all Asian 
countries other than China remained almost constant from 2005 to 2015 and 
as the overall exports increased by 1.6 mbd, it follows that China and India got larger and larger shares of 
the exports from the Middle East production. 

It seems, at this point, that only remarkable geopolitical changes could cause export flows 
from the Middle East OPEC countries to swing away from Asia and towards Western 
Europe and the U.S.

\section{Regional oil production and consumption constraints during the next decades} 

\subsection{The modelled future regional oil production and the results from 2015  and 2016}

The usual method of modelling future oil production for any particular oil field or region is 
limited by how well one knows the volume of extractable oil that remains in the ground, how 
well one understands the technology of extraction that will be used, and how much one knows 
about the future demand for the oil. To date, such knowledge has not been sufficient to 
predict regional oil and oil equivalent production with much confidence. Therefore, another 
way to model future conventional crude oil production has been proposed in Part I of this 
analysis \cite{Part1paper}. The proposed model combines production trends observed 
over the past few years and including the data up to 2014 with the realisation (1) that production 
in all oil fields and regions eventually declines and 
(2) that rate of decline in the larger oil producing regions can be described with an 
approximative universal decline function. In order to obtain predictions for the maximal 
regional oil equivalent production, the production perspectives of unconventional oil,  
NGL's and other oil equivalent liquids including biofuels were largely taken from the EIA perspectives. 
More details about the model and the obtained result can be found 
in the publication \cite{Part1paper}.

\begin{figure}[h]
\begin{center}
\includegraphics[width=13cm]{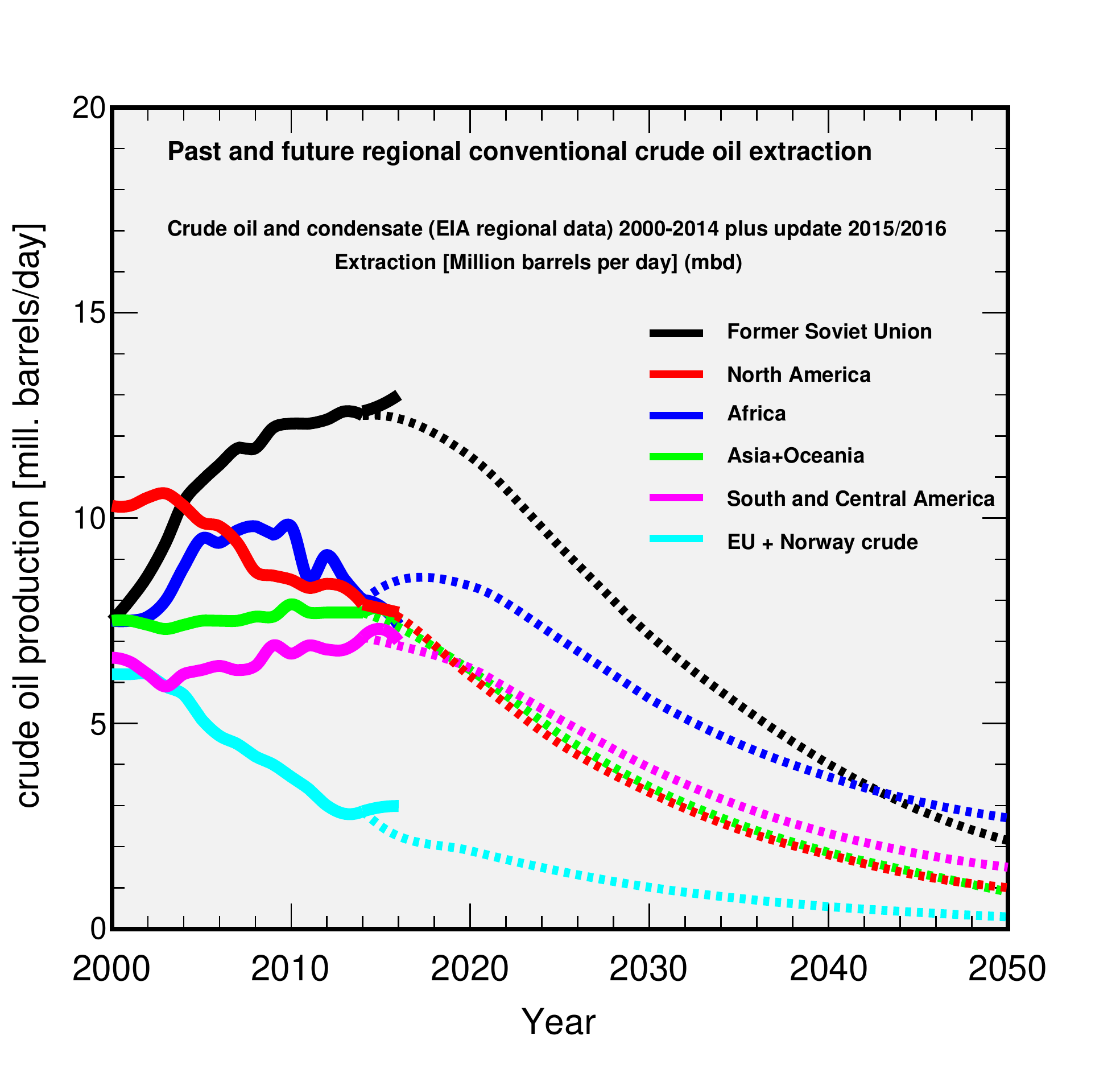}
\end{center}
\caption{\small{The conventional crude oil production data from 2000-2014 (solid lines) were used 
as the input, in Part I of this analysis, to model future production through the year 2050 (dotted 
lines) for all major oil producing regions and continents, outside the Middle East. The latest, 
2015 and 2016 crude and condensate production data, as reported by the EIA \cite{EIA}, are now included in the graph.}}
\end{figure}

The future regional production modelled in Part I, which was based on production data 
through 2014, is shown in the dotted lines in Figure 1. Since the publication of Part I, the 
partially preliminary, regional production data for 2015 and 2016 have become available 
and are now included in the updated Figure 1 as the most recent parts of the solid lines.

When comparing the modelled 2015-2016 production with the 2015/2016 data 
one finds that the production in Western Europe stayed roughly constant, 
while a 6\%/year decline was predicted. However, it should be pointed out that
the observed decline since 2002 in Western Europe was faster than in our model and that the overall 
decline observed between 2002 and 2016 fits so far perfectly to the model. 

In contrast to the modelled ``imminent" beginning of the crude production decline in 
the Western regions of the FSU countries, an overall increased production is reported over the last 2 years (by 3\%). Unfortunately the data have not yet been split into Eastern and Western parts.

The production of conventional crude in the OPEC Middle East countries has increased from 
about 22 mbd in 2010 to about 26.6 mbd in 2016 and is slightly higher but still in agreement   
with Part 1's prediction of a decades-long plateau around 24 $\pm$ 2.5 mbd for this region.
As the OPEC countries and Russia have agreed at the end of 2016 to reduce 
their production slightly \cite{OPECcuts}, in order to increase the oil price, a production decline 
can be expected soon.  

During those 2 years, the production in Mexico declined by 11\% as modelled,   
and a faster than modelled production decline of 7\%, compared to the expected 3\%, 
is reported for China in 2016 \cite{Chinadecline}.  
And finally, the latest production data indicate that South America and Asia are in good agreement 
with the modelled expectations and the production growth in Africa, assumed possible with a  
reduction of the civil war like conditions especially in Libya and Nigeria, has not happened yet. 

Even though one finds a decent agreement, overall, between the modelled production trend
and the regional production data since 2014, it should be clear that natural, economic or 
political forces could cause production fluctuations over any 2-year period that are roughly 
of the same size as the modelled production changes for that 2-year period. Thus the 
real test of the regional production model and its underlying assumptions can probably 
not be given until the data for 2020 become available.

In what follows, the modelled regional production declines (from Part I) and the regional oil
exports trends from section 2 of this paper are combined, such that the regional maximal possible 
oil equivalent liquid energy consumption, and its evolution for the next two decades, can be predicted 
for all major exporting and importing regions and countries.

\subsection{Changes and constraints for future oil exports and imports}

The basic assumptions for the following quantitative prediction for the next 15-20 years of the maximal 
regional oil consumptions are:
\begin{enumerate}
\item The production of oil equivalent liquids will decline, according to the quantitative model described in 
Part I of this analysis, in all regions, outside of the Middle East. 
\item The trend of a modest internal oil consumption growth in oil exporting countries,
as observed during the past decade, will continue for some more years.
\item The future physical oil trade has changed considerable during the last decade, but the  
stability of the now (2015-2017) established export fractions, especially concerning the exports 
from the OPEC Middle East Gulf countries will continue during the next decades.    
\end{enumerate} 

Following those assumptions, the physical flow of oil trade around the globe and 
consequently the regional consumption can be predicted for the next 15-20 years.
Tables 4, 5 and 6 and Figures 2 through 8 summarise the results for production, export and
consumption for the next two decades for all of the major import and export regions of the 
planet and more details will be given below.

{
\begin{table}[h]
\begin{center}
\begin{tabular}{|c|c|c|c|c|c|c|}
\hline
Country/Region & 2015 &  2020 &  2025 & 2030 & 2035 & prod. ratio \\
production upper limit  & [mbd]& [mbd] & [mbd]   & [mbd] & [mbd] & 2035/2015 \\
\hline
U.S. oil equivalent &  12.7    &   12.2   &  11.6 &   10.5  & 10.3  &  0.81 \\
U.S. tight oil only   & 4.6    &     5.      &   5.   &   5.       &   5.   &  1.09 \\ 
EU + Norway        & 3.5     &   2.4     & 1.9   &  1.5       & 1.2   &   0.34\\
China &  4.3    &  3.6    & 2.6  &  1.9  &  1.4  &  0.33\\
India &   0.9    &   0.9   & 0.8  &  0.6   &  0.4  & 0.44\\
Other Asia/Pacific & 3.1 & 2.2   & 1.6  &  1.2  & 0.9 & 0.29  \\
\hline
Middle East &   30.1   & 30 $\pm$ 3 & 30 $\pm$ 3   & 30 $\pm$ 3 & 30 $\pm$ 3 & 1.0  $\pm$ 0.1 \\ 
FSU             &   13.9   & 12.3 & 10.0 & 7.95 &  5.9 & 0.42 \\ 
North Africa  &     2.8    & 3.2 & 2.8   & 2.5 & 2.2 & 0.69 \\
other Africa  &     5.6   & 5.2 & 4.2   & 3.1 & 2.2 & 0.39\\
South America & 7.7 & 7.0 & 6.5   & 5.2 & 4.5 & 0.58\\   
\hline
\end{tabular}\vspace{0.1cm}
\caption{Maximal production of crude oil and other oil equivalent liquids between 2015 and 2035 (the 2015 data are from BP \cite{BPstat}). The difference between the crude production in 2015 and the BP all liquid production, which is roughly up to 10\% larger for different regions is assumed to be constant for the next 20 years and is added to the modelled 2015-2035 crude production estimates.
}
\end{center}
\end{table}
}

{
\begin{table}[h]
\begin{center}
\begin{tabular}{|c|c|c|c|c|c|c|}
\hline
Country/Region & 2015 &  2020 &  2025 & 2030 & 2035 & consumption ratio\\
consumption     & [mbd]& [mbd] & [mbd]   & [mbd] & [mbd] & 2035/2015 \\
\hline
Middle East &   9.6   & 10.5  & 11.0   & 11.0 & 11.0 & 1.15\\ 
FSU           & 4.1 & 4.3 & 4.4 & 4.5 & 4.5 & 1.1 \\
North Africa &  1.3    & 1.4 & 1.4   & 1.4 & 1.4 & 1.08\\
other Africa &   2.6    & 2.9 & 3.0   & 3.0 & 2.2 & 0.85\\
South America &  7.1 & 7.5 & 7.2   & 5.9 & 5.2 & 0.73\\   
\hline 
exports & [mbd]& [mbd] & [mbd]   & [mbd] & [mbd] & 2035/2015 \\
\hline
Middle East &   20.5   & 19.5 $\pm$ 3 & 19 $\pm$ 3   & 19 $\pm$ 3 & 19 $\pm$ 3 & 0.93 $\pm$ 0.15\\ 
FSU             &    9.8      &  8.               & 5.5                 & 3.5             & 1.4               & 0.15\\
North Africa  &  1.4      & 1.8                    &  1.4               & 1.1   & 0.8 & 0.57\\
other Africa &   3.1      & 2.3                    &  1.2               & 0.1   & 0 & 0\\
South America & 1.3   & 0.2                    & 0                   & 0      & 0 & 0\\  
\hline
combined exports &  36         & 32 $\pm$ 3     &  27 $\pm$ 3 & 24 $\pm$ 3 & 21 $\pm$ 3 & 0.59\\
Middle East [fraction] & 57\% & 61\%  & 70\% & 80\% & 90\% & 1.6\\
\hline
\end{tabular}\vspace{0.1cm}
\caption{Consumption and intercontinental crude export evolution in different exporting regions of the planet.
(the 2015 data are from BP \cite{BPstat})
Exports are defined as the differences between production and consumption. The differences 
between production of crude oil, NGL and other oil equivalent liquids and the consumption of 
refined products are assumed to be small enough to be neglected for the accuracy of the modelled predictions. 
}
\end{center}
\end{table}
}

{
\begin{table}[h]
\begin{center}
\begin{tabular}{|c|c|c|c|c|c|c|}
\hline
Country/Region & 2015 &  2020 &  2025 & 2030 & 2035  & 2035/2015 \\
consumption limit        & [mbd]& [mbd] & [mbd]   & [mbd] & [mbd] & ratio \\
\hline
U.S.  (consumption) &  19.4    &   16.5   &  15.6 &   14.5  & 14.3  & 0.75  \\
U.S. (imports total)   & 6.7       &     4      &   4     &     4     & 4       & 0.61             \\
\hline 
EU + Nor + CH (consumption)*& 13.2     &   10.8   & 7.6  &  5.5   & 4.1  & 0.31 \\
EU + Nor + CH (imports)*         &  9.8     &    8.4    & 5.7    &  4     & 2.8 & 0.29  \\
\hline 
China (consumption) &  12.3    &  11    & 9.3  &  8     &  7  &  0.57 \\
China (imports)         &     8.0    &  6.5   & 5.7  &  5.5  & 4.9  & 0.68 \\
\hline 
India (consumption) &   4.2    &   3.9   & 3.3  &  2.9   &  2.7  & 0.64\\
India (imports)         &   3.3     &   3.    & 2.5   &  2.3   &  2.3   & 0.7\\
\hline 
Other Asia/Pacific (consumption) & 11.8 &  10.6 & 10.2  & 9.8  & 9.5  & 0.83 \\
Other Asia/Pacific (imports)         &    8.6 & 8.6   & 8.6  &  8.6  & 8.6   & 1. \\
\hline
\end{tabular}\vspace{0.1cm}
\caption{Maximal consumption and imports of oil equivalent liquids between 2015 and 2035 (the 2015 data are from BP \cite{BPstat}) for the major consumer countries and regions are listed. For the U.S., the imports from Canada (2.5 mbd) and from the Middle East (1.5 mbd) are combined and assumed to be roughly constant during the next decades. *Nor stands for Norway and CH stands for Switzerland.
}
\end{center}
\end{table}
}

However one might have doubts about the third assumption above and it is 
obvious that under a stressed global oil market supply situation, civil war like terrorism or natural disasters within any oil-producing or refining region could quickly reduce production enough to impact total world exports. In addition it does not appear totally imaginary that the competition for the remaining oil 
might increase the conflicts between any of the current world's major powers with catastrophic 
consequences which would invalidate any rational model. In this respect 
it should be obvious for any analyst that the stability in and around the entire Middle East is of 
prime importance for the next decades. As the Middle East is already known to be sensitive to civil wars, resource wars, or any other sort of wars, large or small, it's clear that  
any such event could change significantly the oil production and the export fractions from this region. 

Another theoretically possible Ògame changingÓ development for the oil exports from the Middle East 
might involve Egypt and Turkey. Egypt with its 
growing population already nearing the 100 million mark and Turkey with 80 million people have 
some historical cultural and religious ties to the neighbouring oil rich countries of the Middle East and 
in North Africa. For example, it might be theoretical possible that Egypt's or Turkey's  
leaders might convince those in power in one of the oil-rich states to increase oil exports to Egypt or Turkey
below prevailing world prices as cheap transports could be realised via 1-2 mbd oil pipelines.
The theoretical result could be the creation of a populous and quickly growing economic and political Middle East Union. However, even if some potential ties might remain from the past centuries and during the Arabic and Ottoman empires, it seems that most of them have been successfully replaced by the 
``divide and conquer" colonial policy during the last hundred years. Consequently the 
chances for such a powerful Middle East Union are almost negligible 
during the next decades. As a consequence, neither the peoples in the region 
nor the leaders of Turkey and Egypt should probably not get their hopes up. 
In any case, all such trend changing possibilities are outside of the scope of this analysis which 
tries to determine the maximal possible physical oil consumption during the next 15-20 years. \\

{\bf Production and Exports from the OPEC Middle East countries}\\

\noindent 
Production and interregional export quantities of oil equivalents for the OPEC Middle East
countries are shown in Figure 2. According to the BP reports \cite{BPstat}, 
which document the oil equivalent production, where about 3.9 mbd NGLÕs and some refinery gains are included, the total production increased from 25.6 mbd in 2005 to 28.9 mbd in 2015, and 
to 30.6 in 2016.\footnote{The combined production from other Middle East countries was about 1.2 mbd in 2015 and 2016}. 
 
Most of this increase came from Saudi Arabia, Iraq and Iran with 0.8 mbd, 1.1 mbd and 0.8 mbd respectively. According to the EIA \cite{EIAsparecap}, this production rise is the result of increased use of spare capacity (reducing that spare capacity from about 2 mbd in 2014 to about 1 mbd in 2016), and smaller unplanned 
supply disruptions (which were about 3 mbd in 2014 and only 2 mbd in 2016). 

During the same period, the consumption within the entire Middle East region increased steeply from 
6.6 mbd in 2005 to 9.6 mbd in 2015. As a consequence the intercontinental exports increased only slightly, from 19 mbd in 2005 to 20.5 mbd in 2015, and to 22.4 mbd in 2016.
However, in the fall of 2016, OPEC announced an agreement with Russia and other exporting countries to 
reduce production sufficiently to stabilise the oil price around 50-60 dollars/barrel in 2017, 
OPEC's agreed reduction being about 1.2 mbd \cite{OPECcuts}.
If the announced reduction is achieved, the exports will again be reduced to about 21 mbd, 
which is in agreement with the arguments given in Part I for a rather stable future production of oil
equivalents in the OPEC Middle East region of about 28 $\pm$ 2.5 mbd, and resulting in exports
of 19.0-19.5 $\pm$ 2.5 mbd.


As described in section 2.1, today's physical global oil trade has evolved and changed considerably over recent decades. Especially remarkable is the fact that 16 mbd or 80\% of the OPEC Middle East exports are now flowing to Asian countries and that especially China and India have increased their share of the Middle East oil and imported 3.4 mbd and 2.3 mbd in 2015 respectively\footnote{
The latest BP report indicates that the imports to China and India from the Middle East 
have reached 3.7 mbd and 2.7 mbd in 2016.}.
In contrast Western Europe and the U.S. have significantly reduced their imports from this region
to 1.8 mbd and 1.6 mbd in 2015. 
  
\begin{figure}[h]
\begin{center}
\includegraphics[width=13cm]{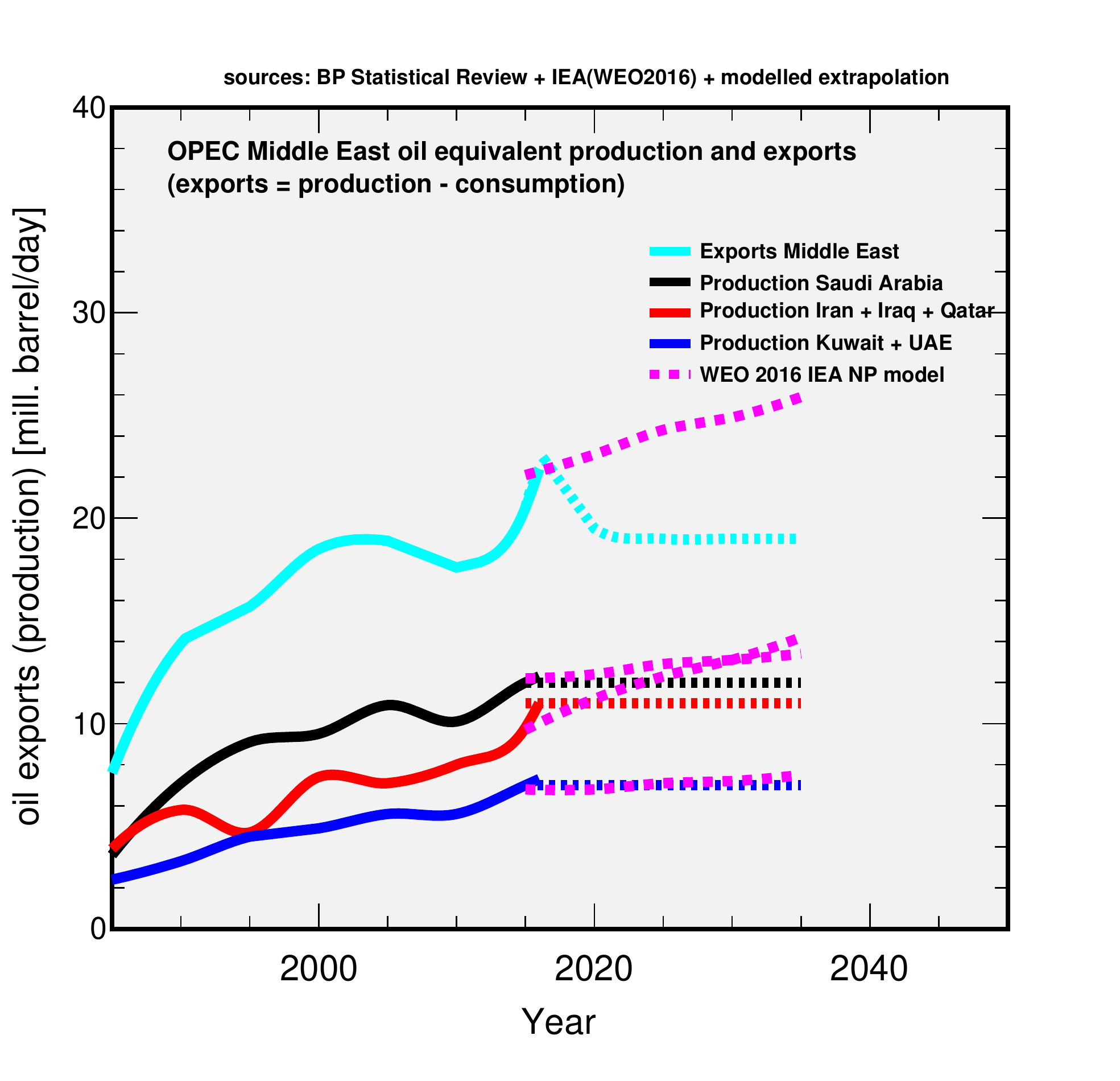}
\end{center}
\caption{\small{OPEC Middle East oil equivalent production and export data through 2016 \cite{BPstat}, 
with production data divided into three groups - Saudi Arabia, Iran + Iraq + Qatar, and Kuwait + 
UAE - is indicated by solid lines. Production and exports through 2035, as modelled in the 
present analysis, are indicated by the dotted lines, while the pink dashed lines indicate the New 
Policies (NP) projections as presented in the IEA World Energy Outlook 2016 (WEO 2016), 
\cite{WEO2016}.
}}
\end{figure}
\begin{figure}[h]
\begin{center}
\includegraphics[width=13cm]{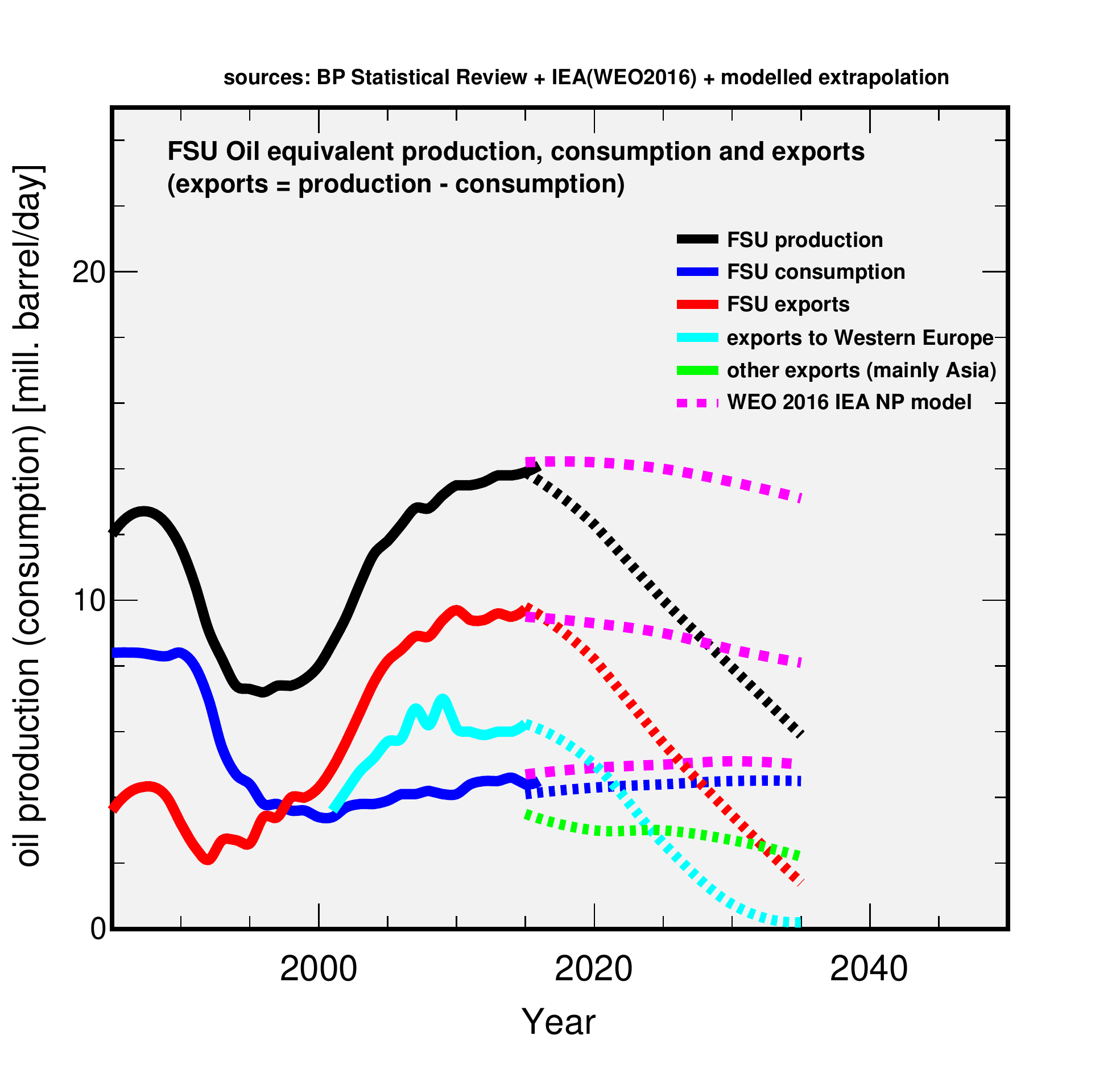}
\end{center}
\caption{\small{Combined oil equivalent production, consumption and exports from three FSU countries, Russia, Kazakhstan and Azerbaijan. Their modelled production, consumption and exports 
following the assumptions from the present analysis, and from the New policies (NP) scenario given in the IEA World Energy Outlook 2016 
(WEO 2016 \cite{WEO2016}), are indicated once again by dotted and dashed lines respectively from 2015 to 2035.
}}
\end{figure}

~~\\
{\bf Production and Exports from Russia, Kazakhstan and Azerbaijan} \\ 

\noindent 
Based on the model presented in Part I, oil production in the FSU countries, and especially 
in Azerbaijan and the Western parts of Russia and Kazakhstan, is predicted to decline by
roughly 3\%/year for 5 years beginning somewhere between 2016 and 2018, and to decline
roughly 6\%/year thereafter.

In contrast, production in the Eastern regions of Siberia and Kazakhstan is expected to 
increase until 2020 and to remain at that higher level until 2025. After that, production is
expected to decline by 3\%/year until 2030, and by 6\%/year thereafter. However, the predicted
production of 2 mbd in Eastern Siberia is expected to flow towards China through pipelines 
that already exist or are under construction.
If that happens as planned, China will, for a time, be able to compensate for the predicted internal production declines.\footnote{For a very recent report 
about the growing Russian oil exports to China see \cite{Russiaoiltochina}.}

If it follows the trend of the last 10 years, internal consumption in the FSU countries will 
continue to increase modestly from its current level of 4.1 mbd to a plateau of about 4.5 
mbd between 2020-2025. Since the oil from the ageing fields in the western regions of 
the FSU countries will have to be shared by Russia, other FSU countries and Western 
Europe, the more oil that Russia and other FSU countries consume, the less will be 
available for export to Western Europe. Hence exports to Western Europe can be 
expected to decline somewhat faster than production declines in the Western parts of 
Russia and Siberia. All things considered, the future exports from the FSU countries to 
Western Europe are estimated to decrease from 6 mbd in 2015 to 4.2 mbd in 2020, 2.2 
mbd in 2025, 1.0 mbd in 2030, and essentially zero thereafter. The past and future oil 
production and export scenarios are summarised in Figure 3.\\

{\bf Production, Consumption and Exports from South America and Africa}\\

\noindent 
The analysis of future exports is further simplified by the assumption that the larger
and richer oil importing OECD countries are not motivated to help the economies of the oil 
producing countries in Africa and South America to grow. This is a 
reasonable assumption because growing economies in those countries would mean greater 
internal oil consumption and hence reduced exports - which is not good for the larger and 
richer oil importing countries.

Thus, following the trends from the last 5 years, the internal oil consumption in oil exporting
countries in Africa and South America can be expected to grow at most 
by about 2\%/year until their declining production matches their internal regional consumption. 
The corresponding consumption growth and the modelled production decline are shown in 
Figure 4 for South America and in Figure 5 for Africa.

Accordingly it can be expected that signiÞcant net exports from South America
to the U.S. and Asia will essentially stop around the year 2020.

\begin{figure}[h]
\begin{center}
\includegraphics[width=13cm]{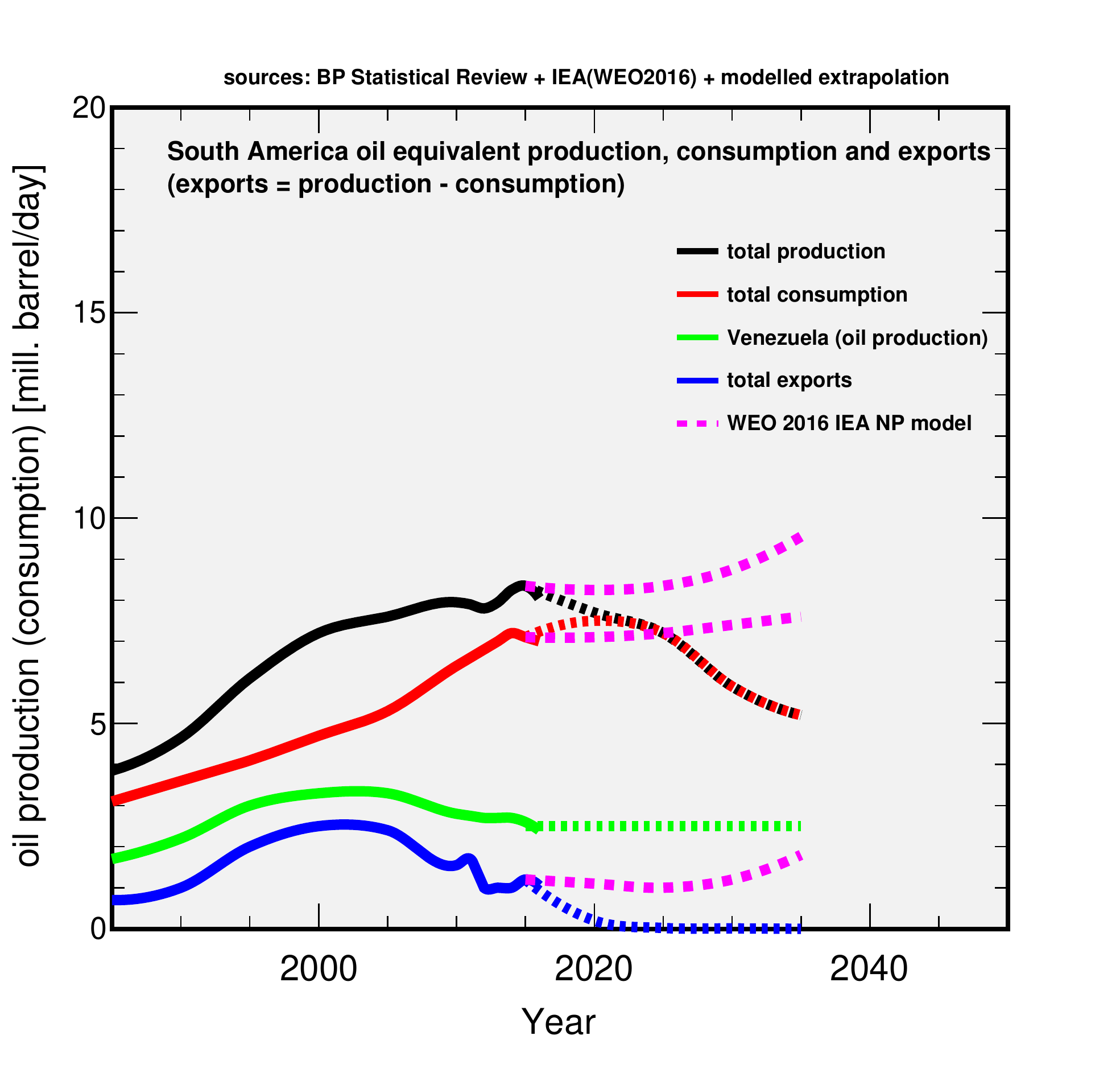}
\end{center}
\caption{\small{South American oil equivalent production, consumption and exports.
Their modelled production, consumption, and exports following the assumptions from the 
present analysis and from the New policies (NP) scenario presented in the IEA World Energy 
Outlook 2016 (WEO 2016 \cite{WEO2016}), 
are indicated by dotted and dashed lines respectively from 2015 to 
2035, just as they were in Figure 2 and 3.
}}
\end{figure}

Due to their current low internal consumption per capita,  or even with moderate growth in that
consumption, exports from Sub-Saharian Africa to Western Europe and other regions
are expected to decline somewhat slower. Those exports are modelled to decrease from the 3.1
mbd (2015) to 2.3 mbd (2020) and 1.2 mbd (2025) and become negligible around the year 2030.

For at least two reasons, the oil production and exports of Northern Africa are difficult to predict:
in Algeria at present, production is declining and consumption is growing, and in Libya there is 
still considerable chaos in the aftermath of the West's bombings and regime change. However, 
if stable production is seen once again in Libya, that country's exports to Western Europe can 
perhaps be reestablished before 2020. And if they are, total exports from Northern Africa to 
Western Europe could increase again to about 1.8 mbd. But after 2020, assuming some growth 
in consumption in Northern Africa, and expecting Algeria's production decline to continue, total 
exports to Western Europe are modelled to decline from 1.8 mbd (2020) to 1.4 mbd and 0.8 
mbd around 2025 and 2035 respectively. \\

\begin{figure}[h]
\begin{center}
\includegraphics[width=13cm]{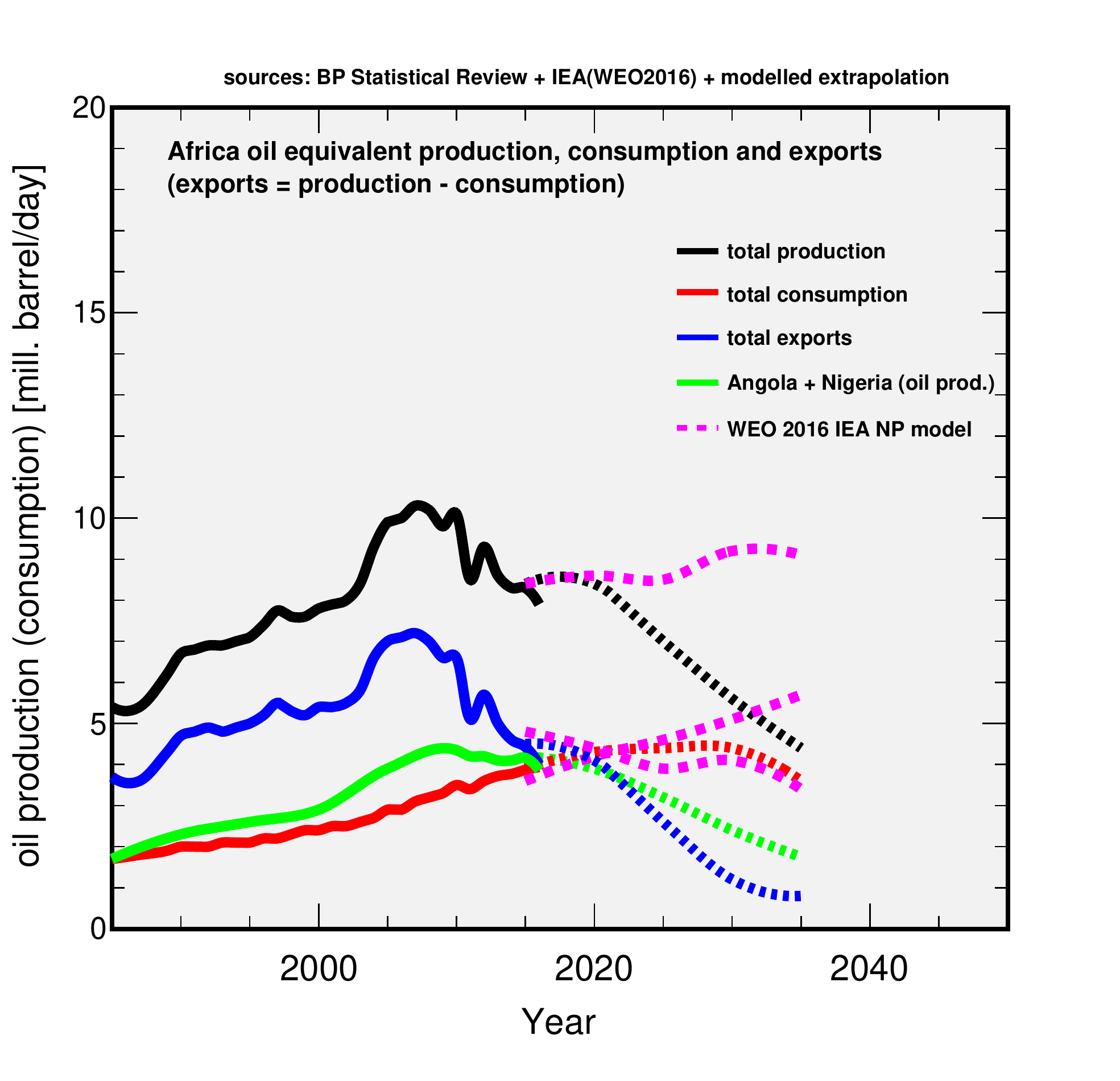}
\end{center}
\caption{\small{Africa's oil equivalent production, consumption and exports.
The modelled production, consumption and exports following the assumptions 
from the present analysis and from the New policies (NP) scenario presented in the IEA World Energy Outlook 2016 (WEO 2016 \cite{WEO2016}), are indicated once again as dotted and dashed lines respectively 
from 2015 to 2035.
}}
\end{figure}

{\bf Production and consumption in Western Europe} \\ 

\noindent 
Since the Western European crude oil production is modelled to continue its average 6\%/year decline, the 
oil equivalent production, which includes about 0.5 mbd from NGL's, are predicted to decline from 3.4 mbd in 
2015 to about 2.7 mbd in 2020, to 2.1 mbd in 2025, and to 1.7 mbd in 2030. Even if one takes the
intermediate plateau like production from 2015 and 2016 into account, the expected production
in 2020 and afterwards might only be about 15\% higher than the numbers given above. 

Except for Norway, every country in Western Europe already depends on imported oil. About
2/3 of the region's imported oil (about 6 mbd) comes from the FSU, and most of it comes 
from Russia. But as explained above, oil production in the FSU countries is expected to decline 
rapidly during the next 10-15 years, and exports from those countries are expected to 
become negligible around 2030. Can Western Europe make up for those declines by importing
oil from elsewhere?

During the past 5-10 years rather constant oil imports of 4-4.5 mbd came to Western Europe
from the OPEC Middle East countries (around 2 mbd), from Libya, from Algeria and, even more,
from West Africa. Following those trends, and assuming a rather constant oil production for the 
OPEC Middle East countries, exports from the OPEC Middle East to Western Europe might 
hold steady around 2.0 mbd. Libya's production might at best reach levels seen before
the chaos by around 2020, and exports from Northern and Western Africa might at best remain 
roughly constant at today's level of 2 mbd and until about 2030.

But for Western Europe to make up for the reduction in imports from the FSU countries it will
be necessary (1) that the African countries will not manage to significantly increase their 
internal consumption of oil and (2) that Western Europe - thanks to politics or shorter shipping 
distances or whatever - will manage to outcompete the U.S., China and other Asian 
countries to get a larger fraction of the African and especially OPEC Middle East's exports.

With no other sources available, the combined physical maximal oil equivalent consumption in Western Europe can now be estimated to decline from 13.2 mbd in 2015 to 
11 mbd (2020), 7.1 mbd (2025), 6.2 mbd (2030) and 5 mbd around 2035.
Those production, import and consumption data and their modelled trends are summarised in Figure 6.   

How this oil will be distributed among the different countries in Western Europe depends strongly 
on the future of the European Union. It might be reasonable to assume that the declining oil exports from Norway, the only oil exporting country in Western Europe, will be shared in fractions similar than today. 
For the oil from the FSU countries, one can expect that most of it will be used within the eastern and more central European countries, including Germany.
Accordingly, the oil imported from Northern Africa will most likely be shipped via the shortest distances directly to Italy, France and Spain. As the oil from the Middle East and Western Africa is shipped with big oil tankers around Africa, it will most likely arrive at the existing big ports like Rotterdam. From there the  
oil will most likely be directly distributed via pipelines to the Benelux region and Germany.


\begin{figure}[h]
\begin{center}
\includegraphics[width=13cm]{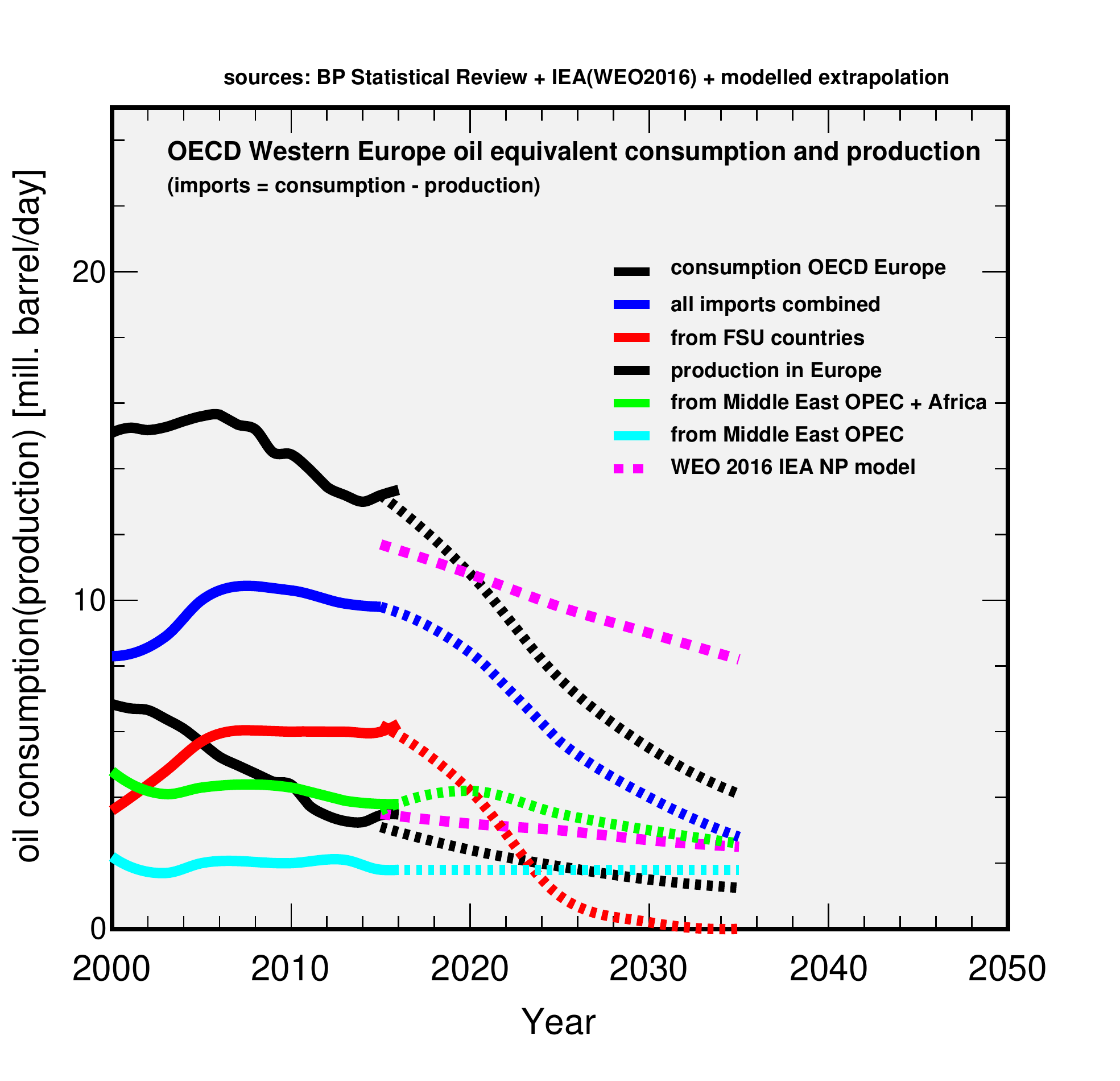}
\end{center}
\caption{\small{Combined oil equivalent production, consumption and imports in Western Europe.
Their modelled production, following the assumptions from this analysis, and from the New policies (NP) scenario given in the IEA World Energy Outlook 2016 (WEO 2016 \cite{WEO2016}), 
are indicated once again by dotted and dashed lines respectively from 2015 to 2035.
}}
\end{figure}

Overall, and following our model with its assumptions, a 50\% reduction of the oil consumption in Western Europe during the next 10-15 years is expected. The decline could happen even faster if  
diplomacy fails to reduce the current tensions with Russia. And any reduction in imports 
from Russia could have more drastic consequences for Germany and the other 
central and eastern countries that receive most of their oil today from Russia.

It appears unlikely that the oil from Algeria and Libya, 
which arrives currently in France and other Southern European countries, 
will for example be easily shared after 2025 with Germany and the other Eastern European countries.   

The economic crisis in the Southern EU countries led to a 20-25\% reduction in their oil 
consumption during the past 5 years. But it can work the other way around too: a reduction
in oil consumption can lead to an economic crisis, which can snowball as less and less oil can
be afforded. And because of that it can be expected that as less and less oil is available 
the so-far-little-affected countries of the EU and Switzerland (CH) will enter into a continuing 
economic and political crisis period, with effects similar to what was observed in Greece 
during the last 5 years. The car making companies can be expected to lead the way down.

Summarising the above, it seems that the people in the Western European countries will 
soon be faced with an unavoidable downward spiral in their oil based economies. It is 
expected that this crisis will begin no later than 2020 when the oil production in the 
Western parts of Russia begins its final decline. \\

{\bf Production and consumption in North America and in the U.S.} \\ 

\noindent 
For the U.S., the expected evolution of the oil equivalent production, consisting of conventional oil, NGL and the tight oil contribution is quantified below and shown in Figure 7, as is the U.S.
consumption.

\begin{itemize}
\item Conventional oil production is predicted to decline from 4.3 mbd (2015) to about 3.7 mbd (2020) and 2 mbd (2030).  
\item Tight oil production, NGL and other oil equivalent liquids might follow the optimistic predictions from the EIA, and if they do they will increase from about 8 mbd in 2015 
to a combined plateau of 9 mbd between 2020 and 2030.
\item The exports from the Middle East, 1.5 mbd, and Canada, 2.5 mbd, will remain at this level until about 2030. 
\item All other imports, including the ones from Mexico, will continue the decline observed during the last 10 years. Similarly, significant imports from South America and West-Africa are expected to 
end by 2020. However, the U.S. might be able to win an increasing share of the 2 mbd 
of oil produced by and exported from West-Africa. If that happens, perhaps as much as 1 
mbd could be shipped to the U.S. instead of Western Europe, but this would certainly 
increase the tensions between the U.S. and Western Europe.
 \end{itemize}  

\begin{figure}[h]
\begin{center}
\includegraphics[width=13cm]{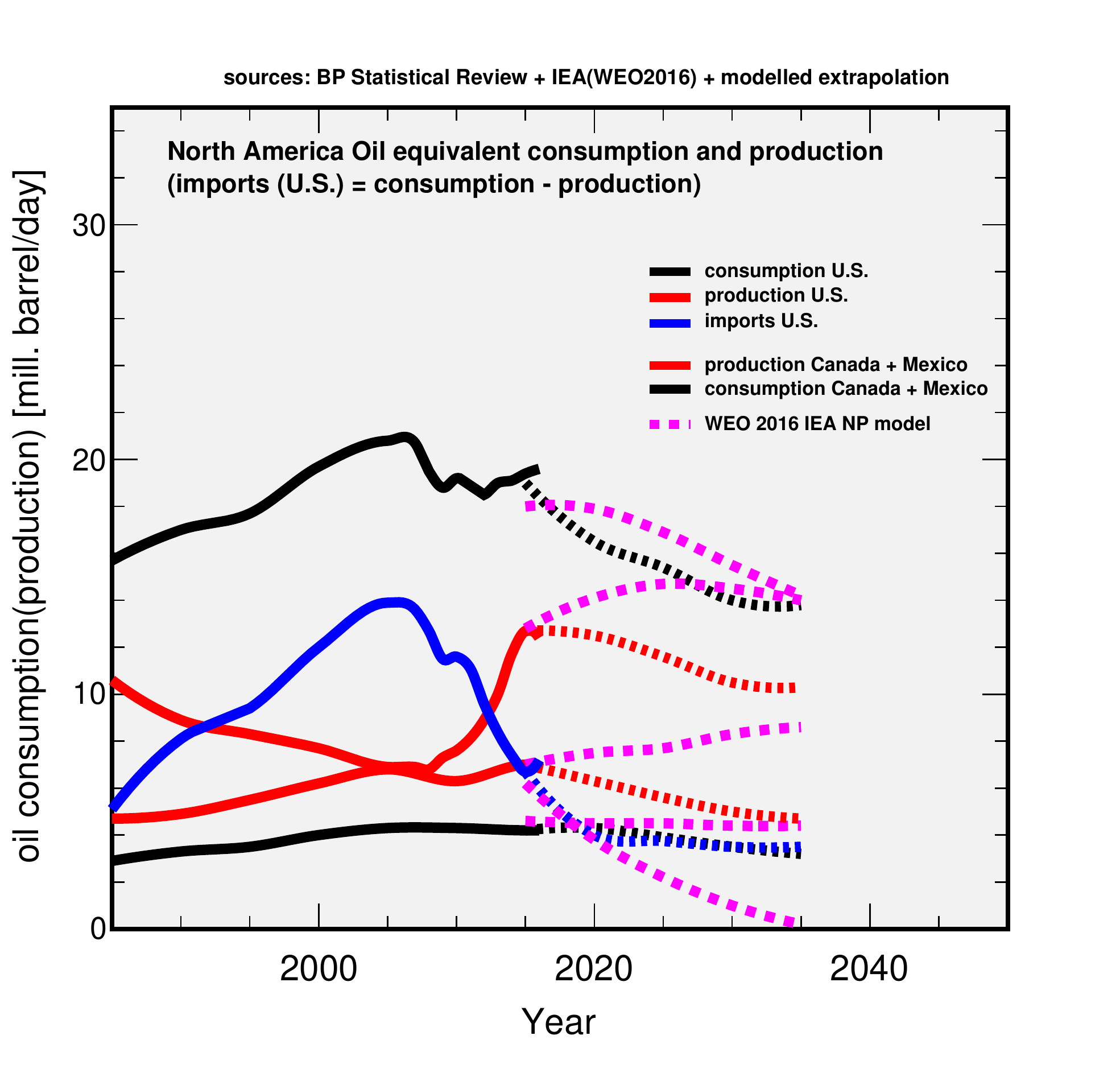}
\end{center}
\caption{\small{Oil equivalent production, consumption in North America and U.S. imports. 
Their modelled production and consumption, following the assumptions from this analysis, and from the New policies (NP) scenario given in the IEA World Energy Outlook 2016 (WEO 2016 \cite{WEO2016}), 
are indicated once again by dotted and dashed lines respectively from 2015 to 2035.
}}
\end{figure}

When adding those numbers, the internal maximum possible consumption in the U.S. is expected to decline from 19.4 mbd in 2015 to 16.5 mbd by 2020 (about 5\%/year), followed by a somewhat slower decline 
between 2020-2030. Specifically, overall consumption in the U.S.
is predicted to decrease to 15.6 mbd in 2025 and to 14.5 mbd in 2030. However, this slower 
decline will not be seen unless the tight oil production decline observed during 2016 
can be reversed so tight oil production can be increased during the years just ahead to about 5-6 mbd.
~~\\

{\bf Production, Imports and consumption for the Asian Pacific region} \\

\noindent 
For China and also for all Asian countries combined, production is expected to decline by
about 6\%/year after 2020. In addition, the competition with the U.S. and Western Europe
for the declining production of Africa and South America, which were exporting about 1.7 mbd
and 1.4 mbd respectively to China and India combined in 2015, will likely lead to the end of 
significant exports to Asia by around the 2020. Thanks to its energy relationship with Russia, 
and if the new fields in Eastern Siberia and Kazakhstan grow as planned, 
China might soon be able to import about 2 mbd of that oil at least through 2030.
China is thus perhaps in a better position than other Asian countries and may be able to compensate for 
the loss from the decline of internal oil production and for some loss from the declining 
oil production and exports from Africa and South America. 

As explained, one might expect that the oil exports from the OPEC Middle East countries to 
Asia will be rather constant during the next couple of decades. But the future oil consumption 
of the various different Asian countries will depend on the respective shares of that oil that 
go to each - China, Japan, India, and South Korea in particular. If their respective shares 
remain fairly constant, and if they collectively import no more from the OPEC Middle East 
than they do now, then about 20\% of OPEC Middle East oil will continue to be available for
export to Europe and the U.S.

The predicted future production, imports and consumption for most Asian countries are
shown in Figures 8 and 9. As shown, the rapid growth of oil consumption 
in China and India - often more than 5\%/year during the last decade - is predicted to come 
to a rather sudden end before 2020. And since China's own production of oil appears to 
have begun to decline in 2016, one can predict that Chinese oil consumption will soon begin
to decline as fast as its internal oil production, which is to say, by about 1 mbd every 5 years
for the next 10-15 years.

Future oil consumption in Japan, South Korea, Taiwan and Singapore will be based almost entirely on imports from the OPEC Middle East countries, since their small imports from 
other Asian countries will come to an end over the next 2 to 3 years.

For the remaining countries in Asia and Oceania, which have little or no likelihood of
establishing new ties with the OPEC Middle East countries, it can be predicted that 
their consumption of oil will be limited to what they can produce, and that what they can
produce will decline steadily.

\begin{figure}[h]
\begin{center}
\includegraphics[width=13cm]{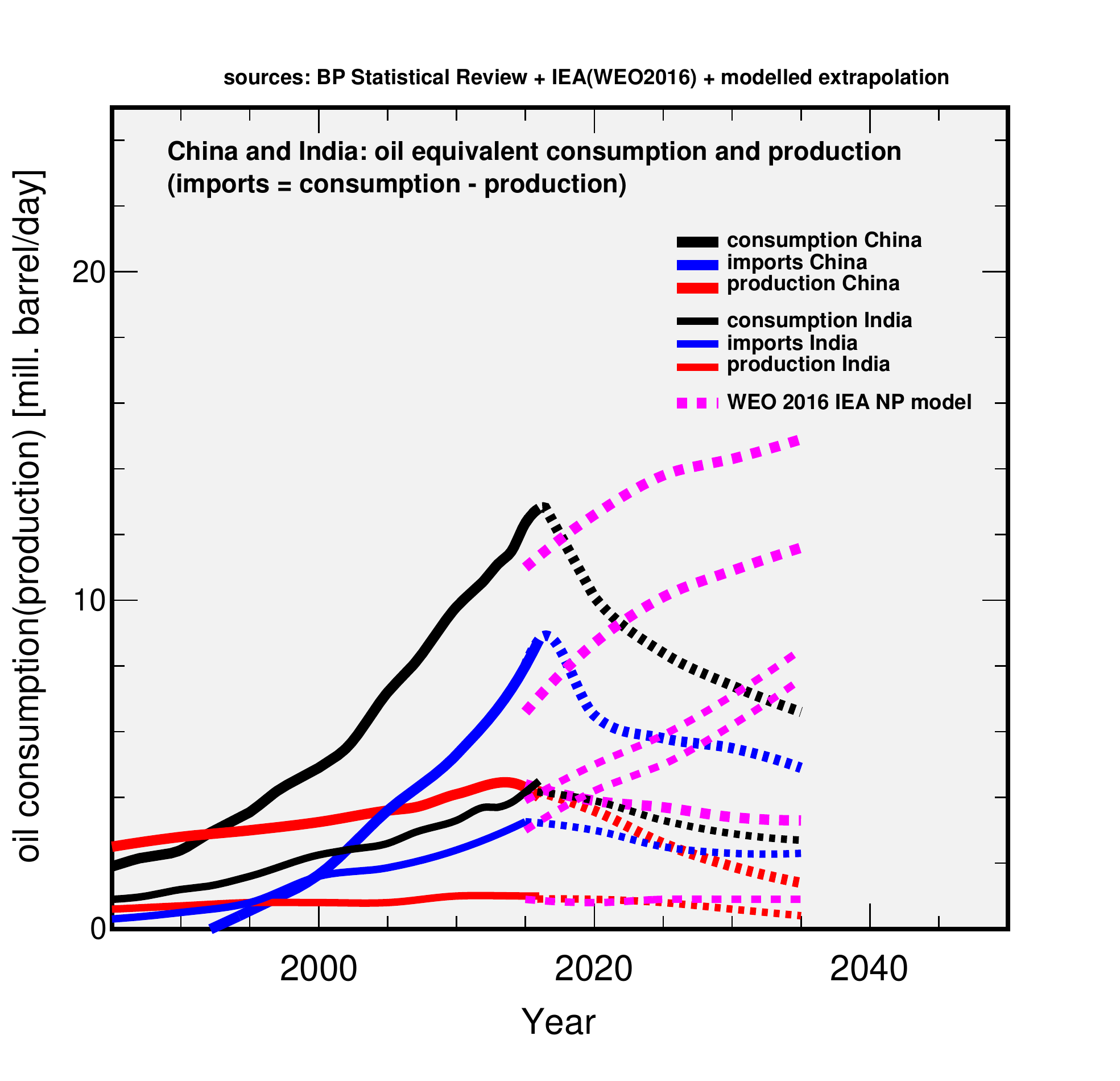}
\end{center}
\caption{\small{Combined oil equivalent production, consumption and imports for China and India. The 
modelled predictions, following the assumptions from this analysis, and the predictions from the New policies (NP) scenario presented in the IEA World 
Energy Outlook 2016 (WEO 2016 \cite{WEO2016}), are indicated once again by dotted and dashed lines 
respectively from 2015 to 2035.
}}
\end{figure}

\begin{figure}[h]
\begin{center}
\includegraphics[width=13cm]{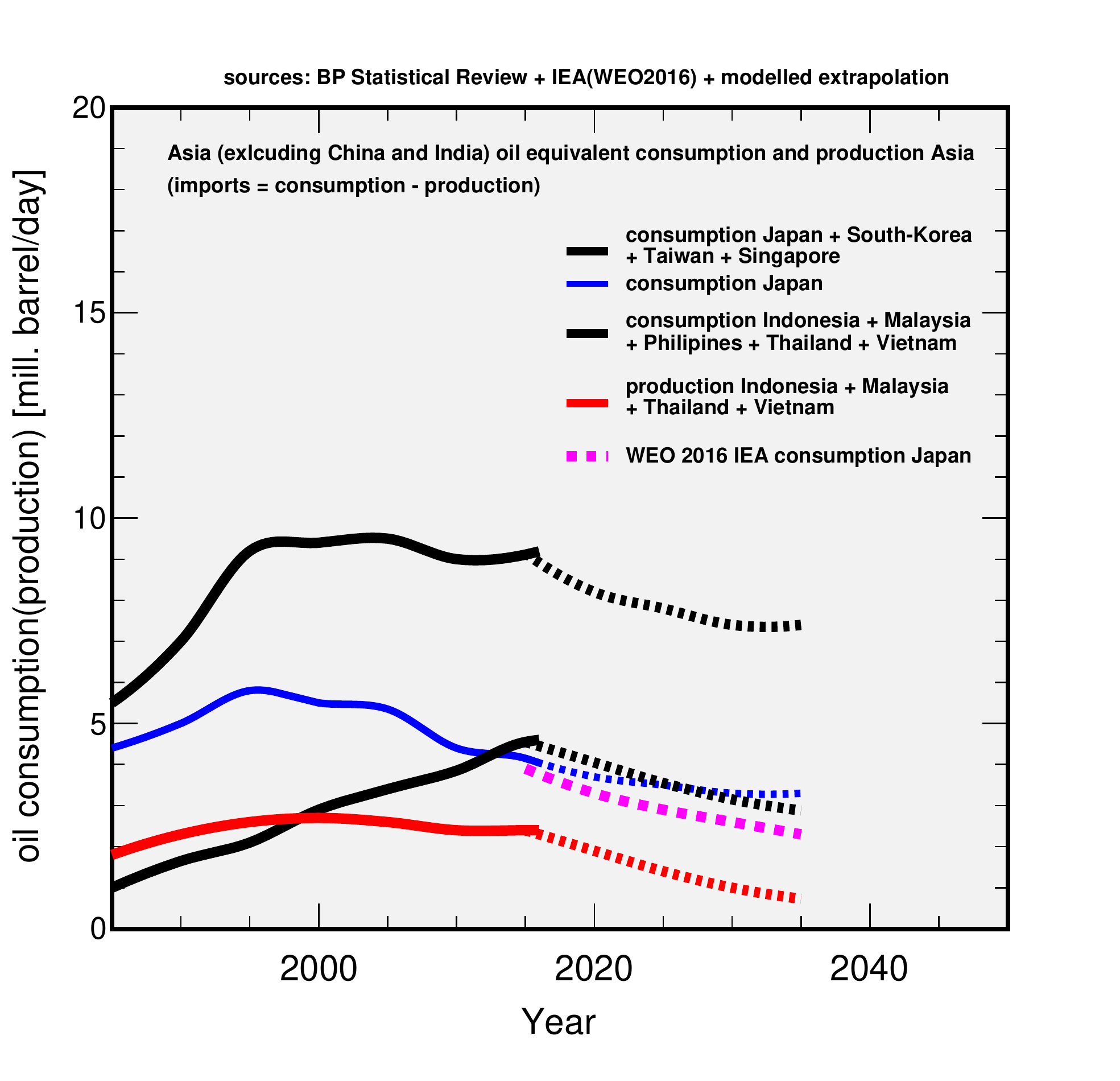}
\end{center}
\caption{\small{Combined oil equivalent production, consumption and exports for the Asian region, excluding
China and India.The modelled predictions, following the assumptions from this analysis, are indicated by dotted lines. The prediction for Japan from the New Policies (NP) scenario presented in the IEA World Energy Outlook 2016 (WEO 2016 \cite{WEO2016}), is indicated by a dashed line from 2015 to 2035.
}}
\end{figure}

\section{Comparing this study's regional and global oil production and consumption predictions 
for the next two decades with the IEA's  
economic-development-based predictions} 

Previous sections have presented the recent-decades evolution of regional oil production, and of 
intercontinental oil trade and consumption. They have also presented this study's modelled 
predictions for the period 2015 to 2035, predictions that assume there will not be so much
fighting or terrorism in the OPEC Middle East countries that it will disrupt their rather stable 
exports of around 19 to 20 mbd. The results are summarised in Tables 4, 5 and 6 above. 
Depending on the region, the annual oil consumption decline in essentially every oil importing 
country is estimated to be at least 2 to 5\%/year in the near future and especially after 2020. \\

This section will explore the deep disagreements between that prediction and the more 
optimistic global and regional oil demand scenarios that are presented in the annual international
energy outlooks published by organisations like the IEA \cite{WEO2016} and the EIA \cite{EIAfuture}.

In particular, the results obtained in the present analysis are contrasted with the predictions
from the IEA WEO 2016 New Policies (NP) scenario for the years 2020 through 2035 \footnote{Quote ``The New Policies Scenario, the central scenario in this World Energy Outlook (WEO) incorporates all policies and 
measures that are in place today, while taking into account, in full or in part, the aims, targets and intentions
that have been announced, even if these are not fully implemented."\cite{WEO2016NP}}. 

Before examining the differences, however, it should be noted (a) that global production 
always satisfied global demand in the IEA scenario and (b) that the oil consumption in 2015 was significantly higher in the oil importing OECD regions of Western Europe, Asia and North America and also in the Middle East region, when compared to the demand assumed within the NP scenario. 
In particular the oil consumption in the OECD countries in Europe is reported as 13.2 mbd 
while the oil demand (consumption) in the NP scenario was assumed to be 11.7 mbd, and 
to decline from there to 10.8 mbd in 2020 and to 8.2 mbd in 2035.
A similar discrepancy for 2015 is seen for the U.S., where the consumption was expected 
to be around 18 mbd, while the actual 2015 consumption is reported to have been 19.4 mbd. 

As the IEA tried to include some expressed government policies from past years, 
presumably envisaged to reduce CO2 emissions, it might be concluded that either the 
government policies or the underlying demand assumptions in this NP scenario are already known 
to be unrealistic.  \\

Concerning the future regional production, the underlying assumption in the NP scenario
seems to be that the declining production in the ageing oil fields in most non OPEC regions 
around the planet is a real problem. But they assume this 
unavoidable production decline can be reduced for example 
to about 1.8\%/year in Western Europe, 1.1\%/year in Russia and 1.3\%/year in China.  
And they seem to assume further that the steep decline from old fields can be compensated, at least up to 2040, with:
\begin{itemize}
\item the discovery and exploitation of new fields in those producing countries and regions and -thanks 
to technology - higher ``technically recoverable resources" in the old and new fields, 
\item new technologies that will reduce the demand for oil equivalent liquids in OECD countries 
and other technological advanced countries,
\item better profitability of investments in oil production as oil price rise, and 
\item market forces that will be sufficient to stimulate the OPEC Middle East producers 
to increase production enough to accommodate demand growth in China and India. 
\end{itemize} 
     
Perhaps the most significant difference between the predicted 6\%/year decline (after 2020) 
in this analysis and the WEO 2016 NP scenario, is that in the IEA view, much smaller regional oil production declines of 1.3-1.8\%/year are assumed for Western Europe, Russia and China (more details are given in Table 7). In addition, the NP scenario assumes that the OPEC Middle East oil production will increase from 29 mbd (2015) to 35 mbd (2035) and about 2/3 of this increase is supposed to come from Iraq and Iran, and
most of this additional oil will be exported to satisfy the growing demand especially in Asia. 
In addition, they assume that the additional global demand growth will be satisfied 
with oil equivalent liquids, like NGL's and biofuels which are assumed to increase 
from 16 mbd (2015) to 21 mbd (2035) and from 1.6 mbd (2015) to 3.6 mbd (2035) respectively. 

More details from the NP scenario in the WEO 2016 report are presented in Table 7 and 
some of the predicted trends, where direct comparisons are available are included in Figures 2-9.
Obviously, the expected consumption trends during the next 10-20 years are totally different 
in the model presented in this paper. 

So perhaps the best test of the production-decline model presented in the present 
study versus the optimistic, growth-assuming, economic-demand scenario 
presented by WEO 2016 are:

\begin{itemize}
\item The present study's predicted oil production declines 
in Western Europe, in China and in the Western Parts of Russia and Kazakhstan
between now and 2020, and between 2020 and 2025, and 
\item the future production from the OPEC Middle East countries and the shares of their exports  
that will go to China and India on the one hand and to the U.S. and the OECD countries in Western Europe and Asia on the other.
\end{itemize}
 
{
\begin{table}[h]
\begin{center}
\begin{tabular}{|c|c|c|c|c|c|c|}
\hline 
 
                               & BP data & \multicolumn{5}{|c|}{IEA WEO 2016 NP scenario} \\ 
Country/Region                     & 2015/2016       & 2015 &  2020 &  2025 & 2030 & 2035 \\
                                              & [mbd]         & [mbd]& [mbd] & [mbd]   & [mbd] & [mbd]  \\ \hline
                                              & \multicolumn{6}{|c|}{demand (=consumption)} \\ \hline
United States                        & 19.4/19.6    & 18.0 &   17.9   &  16.9 &  15.5  & 14.2    \\
Asia Oceania (OECD)          &    7.9/8.0      &    7.2 &     6.6   &  6.0 &   5.5  & 5.1    \\
OECD Europe        & 13.2(+0.8)/13.4(+0.9)& 11.7 &   10.8   &  9.8 &   9.0  & 8.2    \\ \hline 
China                                   & 12.3 /12.8            & 11.0 &   12.6   &  13.8 &  14.3  & 14.9    \\ 
India                                     &  4.2/4.5             &     3.9 &     5.0   &  5.9 &      7.1  &  8.5    \\ \hline
Middle East                          &   9.6(3)/9.4             &    7.9 &   8.5   &  9.2 &  9.7  & 10.3    \\
Russia                                  &   3.1/3.2    &   3.0 &   3.0   &  3.1 &   3.1  & 3.0    \\  
Africa                                    &   3.9/3.9    &   3.6 &   4.2   &  4.6 &   5.1  & 5.7    \\
Latin America                       &   7.1/7.0    &   5.8 &   5.8   &  5.9 &  6.1  & 6.3    \\ 
World total liquids                &  95.0/96.6   &   94.1 &  97.9   &  100.8 &  102.8  & 105.3    \\
\hline 
& \multicolumn{6}{|c|}{production} \\ \hline
United States                   &   12.7/12.4           &  12.8   &    14.1      &   14.7 &  14.5    & 14.0    \\  
OECD Europe                  &     3.4/3.5             &   3.5   &     3.2      &   3.0 &  2.7    & 2.5    \\     
China                               &     4.3/4.0             &   4.4   &     3.9      &   3.7 &  3.4    & 3.3    \\
India                                 &      0.9/0.9          &  0.9     &     0.8      &   0.9 &  0.9    & 0.9    \\
\hline 
OPEC Middle East       & 28.9/30.5    & 28.7 &   30.4   &  32.3 &  33.4  & 35        \\
other OPEC countries &   9.3/8.8    & 10.4 &   8.8   &  9.0 &  9.5  & 10.1    \\
Russia                         &  11.0/11.2  &  11.1 &   10.9   &  10.5 &   9.8   & 9.3    \\
Asia/Oceania              &     8.3/8.0   &   8.0  &    7.2   &  6.8 &   6.5   & 6.3    \\
Africa                          &    8.4/7.9    &   8.4  &   8.6   &  8.5 &   9.2   & 9.1    \\
Latin America             &    7.7/7.5    &  7.7 &     7.6   &  7.6 &   8.0   & 8.8    \\

\hline
\end{tabular}\vspace{0.1cm}
\caption{The oil demand (= consumption) and production numbers as assumed in the 
New Policy scenario from the 2016 IEA World Energy outlook \cite{WEO2016}  
for selected countries and regions. Turkey, with a 2015 oil consumption of 0.8 mbd is 
included in the IEA numbers for OECD Europe.   
For comparison the actual 2015/2016 production and consumption data from BP \cite{BPstat} 
are also given.   
}
\end{center}
\end{table}
}

\section{Summary and Conclusions}

Oil became the energy source of choice during the second half of the 20th century. While 
other energy sources were found suitable for heating and electric power generation, the 
advantages of a liquid fuel for the transport of people, consumer goods and military materials, 
by road, ship and air, were overwhelming. As a result, oil has come to be seen as the very 
"blood" of the modern world's economies. Even modern agriculture depends on it, and not 
only to power large farm machinery but also as the feedstock for fertilisers and pesticides and 
especially for the transport and distribution of the agricultural products from the countryside 
to the cities. 

Modern oil-based societies with their many comforts and luxuries are, however, enjoyed by 
at most 10 to 15\% of the world's people. In 2015, for example, 7.4 billion of us used about 
94 million barrels of oil and oil equivalents a day worldwide, about 2 liters per capita per day 
on average, but only about 1/7 of the people on the planet are significant consumers of oil. 
The average consumption per day in the European Union is about 5 liters a day. If it takes 
that much oil to enjoy a good, comfortable lifestyle these days, 6/7 of us are not even close.

Most of the industrialised OECD countries are now highly dependent on imported oil, 
Norway and Canada being the exceptions. For most of the past 50 years, the major OECD 
oil importers were the only countries that were making things the oil exporting countries 
wanted to buy in exchange - a variety of consumer goods, vehicles, and weapons, and of 
course a lot of luxury items in the OPEC Middle East. But for more than a decade now 
China has become a dominant exporter of many of things that the oil exporting countries 
want, and more recently India has as well, and that has led both China and India to become 
competitive oil importers. Both are taking growing fractions of available global exports.

Based on recent oil production trends in all regions of the planet, a model for future oil 
production for all regions was developed in Part I of this analysis. That model revealed 
maximal production levels, over time, as total production inexorably declines. 

In this paper, Part II, oil consumption trends, combined with the production trends from
Part I, provide a model for future export trends. Combining the three - production, 
consumption and exports - future regional oil consumption for all regions of the planet
are modelled.

Aside from the OPEC Middle East region, where a rather stable production is modelled 
for the next 15 to 20 years, production in essentially all other regions is predicted to be
declining by 3 to 5\% per year after 2020, and some are already declining at this rate. 
Based on the evolution of intercontinental oil exports during the past decade, it is 
predicted that in the near future Western Europe will not be able to replace steeply 
declining exports from the FSU countries, and especially from Russia. Hence total 
consumption in Western Europe is predicted to be about 20\% lower in 2020 than it was 
in 2015. For similar reasons, although the export sources are different, total consumption 
in the U.S. is predicted to be about 10\% lower. 

Further, it is predicted that neither India nor China will be able to continue their rapid oil 
consumption growth. At best both countries might be able to stabilise per capita oil
consumption close to their current relatively low levels through 2025 by outcompeting other countries 
in Asia. 

To put it mildly, the obtained modelled results for future regional oil consumption in almost 
every part of the planet disagree strongly with essentially all economic-growth-based 
scenarios like the one from the IEA in their latest WEO 2016 report. Such scenarios 
assume ongoing growth and would have us believe that the oil required to support such 
growth will be discovered and produced. It won't. Even if the models presented in Part I 
and Part II of this analysis are not perfect, they do reflect the Limits to Growth that are at 
this point becoming more obvious by the day. By 2020 it may be clear to almost everyone 
that the current oil-based way of life in the developed and developing countries has begun 
a terminal decline. 

Whenever that terminal decline begins, one can only hope that people around the globe 
will be able to learn, quickly, how to live with less and less oil every year, and how to 
avoid war and other forms of violence, as we travel the path to a future with less and less oil.

In order to conclude this article on a more positive note, let us hope that the end of the oil era,
and indeed the end of the fossil-fuel era, will lead not only to reduced CO2 emissions 
and less Global Warming than would otherwise have been seen, but to new, locally-oriented, 
more conscious and more ethical ways of life. Let us hope that each of us can 
become committed to leaving the worst aspects of the 20th century in history's dustbin.\\

~~\\
\noindent

{\bf \large Acknowledgments\\} 

{\normalsize  \it{While the ideas presented are from the author alone, my thanks go to 
my friends, colleagues, and many students who have helped me during extensive exchanges to 
formulate these points and arguments. Especially I would like to thank W. Tamblyn for almost uncountable 
exchanges and discussions about regional and global energy and un-sustainability problems since more than 10 years which helped me to formulate the ideas presented here. I would also like to thank him especially for the patience with his careful and critical reading of several versions of the evolving manuscript and for the many suggestions on how to improve the presentation of the ideas in this paper. 
I would also like to thank T. Ruggles, who recently became interested in the model, for several interesting discussions and his helpful suggestions to improve the clarity of the arguments presented in is paper. 
}}



\begin{thebibliography}{99}
\bibitem{EIAoilhistory}{The oil and energy consumption data for most countries between 1965 and 2016, according to the BP Statistical Review of World Energy, can be found at \url{http://www.bp.com/en/global/corporate/energy-economics/statistical-review-of-world-energy.html}.
More data and especially the historic data for the U.S. 
from 1775-2011 for the oil and energy consumption in the U.S. can be found 
in the Total Energy annual report from 2011 at 
\url{https://www.eia.gov/totalenergy/data/annual/pdf/perspectives_2011.pdf}.
For a review about GDP statistics and the estimate of historic GDP/capita growth in different countries 
see Prof. D. Coyle and especially Figure 1 in \url{https://www.uschamberfoundation.org/article/why-gdp-statistics-are-failing-us}. 
}  
\bibitem{CIAfact}{Population data for the different countries were taken from the CIA World Fact Book,
\url{https://www.cia.gov/library/publications/the-world-factbook/}.} 
\bibitem{EIA}{Data on crude oil and oil equivalent production and reserves were taken from 
the U.S. Energy Information Agency (EIA) \url{https://www.eia.gov/petroleum/}.} 
\bibitem{BPstat}{Most oil equivalent production, consumption and global trading data 
were taken from different annual Statistical-Review-of-World-Energy reports, provided by BP on an 
annual basis since many decades. The June 2017 review with the 2016 data can be found at 
\url{http://www.bp.com/en/global/corporate/energy-economics/statistical-review-of-world-energy.html}
and the June 2016 review with the 2015 data at \url{https://www.bp.com/content/dam/bp/pdf/energy-economics/statistical-review-2016/bp-statistical-review-of-world-energy-2016-full-report.pdf}.}
\bibitem{Part1paper}{M. Dittmar, 2016: Regional Oil Extraction and Consumption: A Simple Production Model for the Next 35 years Part I.  BioPhysical Economics and Resources Quality 1:7 
DOI 10.1007/s41247-016-0007-7.}
\bibitem{OPECcuts}{For the 2016 OPEC agreement and the recent announcements that indicate that this production maximum will be extended see: \\
\url{https://www.opec.org/opec_web/static_files_project/media/downloads/press_room/OPEC\%20agreement.pdf} and
\url{http://www.marketwatch.com/story/oil-prices-steady-with-all-eyes-on-opec-meeting-2017-07-24}.} 
\bibitem{Chinadecline}{According to the 2017 BP report, \cite{BPstat}, the 2016 production in China 
was about 4mbd, a reduction of 0.3 mbd compared to 2015. The preliminary data  
indicate that the production in May 2017 has further decreased to an average of 3.83 mbd, 
\url{http://www.reuters.com/article/us-china-economy-output-crude-idUSKBN1950C2}.} 
\bibitem{EIAsparecap}{For more details see the EIA report (most recent July 2017) about the OPEC oil supply situation at \url{https://www.eia.gov/finance/markets/crudeoil/supply-opec.php}.}
\bibitem{Russiaoiltochina}{The June 2017 oil import data from China indicate that Russian exports  
to China have so far increased by almost 30\% compared to 2016 \url{https://www.reuters.com/article/us-china-economy-trade-oil-idUSKBN1A90QY}.} 
\bibitem{WEO2016}{The most recent IEA World Energy Outlook 2016 and the ones from previous years can be found at the IEA website at \url{http://www.worldenergyoutlook.org}.} 
\bibitem{EIAfuture}{Similar to the IEA World Energy Outlook, the EIA provides 
an annual International Energy Outlook report which can be found 
at \url{http://www.eia.gov/forecasts/ieo/}.}
\bibitem{WEO2016NP}{ The quote is from section 5.2.2 of the WEO 2016 \cite{WEO2016}.}





\end{thebibliography}
\end{document}

It was also concluded that, without military adventures, the current exports of roughly 20 mbd will continue during the next decades and according to todays established export lines. 
Under these conditions one can expect that about 16 mbd (80\%) of this oil will be exported to the different Asian countries. Accordingly, the remaining 4 mbd (20\%) will be exported in roughly equal fractions to the U.S. and Western Europe. However, following the arguments from part I of this analysis, a significant decline 
of the oil production is modelled for essentially all other exporting is expected around the year 2020, and essentially only two options remain especially Western Europe, the U.S. and the bigger Asian countries.  

Either they find a way to (1) convince the OPEC Middle East countries to increase their oil production, 
and export in exchange for whatever "monetary price", which will increase the global 
influence of the Middle East OPEC countries, or  (2) by trying to change the established export fractions with the help of a neocolonial occupation of this oil rich Gulf region.

If one takes the trends of the past decade, it seems that especially the economic exchange, 
oil against products,  and the related politics have allowed China and India to outcompete 
the OECD countries as they are getting more and more access to this Middle East OPEC oil. 
Furthermore, knowing that the thirst for an ever increasing oil consumption in China and India  
is still enormous an increased oil production option will most likely only benefit the Asian countries.  
It thus appears unlikely that the reversal of the recent oil trading trends will be reversed in favour of 
Western Europe or the U.S. and that this can be achieved without war like interventions against the competition from China and India.